\documentclass[10pt,a4paper,twocolumn]{article}
\AtBeginDocument{}
\usepackage[linkcolor=blue, citecolor=blue, urlcolor=blue, colorlinks=true]{hyperref}

\usepackage{graphicx}
\usepackage{amsmath}
\usepackage{amsfonts}
\usepackage{amssymb}
\usepackage{bm}
\usepackage{ulem}
\usepackage[usenames, dvipsnames]{color}
\usepackage{booktabs}
\usepackage{soul}
\usepackage{afterpage}
\usepackage{float}
\usepackage{LKBlogo} 
\usepackage{cite}
\usepackage{balance}
\usepackage[sectionbib]{bibunits} 
\usepackage{caption}
\DeclareMathOperator\Arg{Arg}

\newcommand{\phir}{\phi_{\mathrm{r}}}
\newcommand{\phid}{\phi_{\mathrm{d}}}
\newcommand{\phis}{\phi_{\mathrm{s}}}
\newcommand{\phie}{\phi_{\mathrm{e}}}
\newcommand{\ts}{t_{\mathrm{s}}}
\newcommand{\tp}{t_{\mathrm{p}}}
\newcommand{\td}{t_{\mathrm{d}}}
\newlength{\bibitemsep}
\newlength{\bibparskip}
\let\oldthebibliography\thebibliography
\renewcommand\thebibliography[1]{  \oldthebibliography{#1}  \setlength{\parskip}{\bibitemsep}  \setlength{\itemsep}{\bibparskip}}

\begin{document}
\defaultbibliographystyle{elsarticle-num} 
\defaultbibliography{BSpaper}

\title{Generating accurate tip angles for NMR outside the rotating-wave approximation}
\shorttitle{Bidinosti et al, NMR outside the rotating-wave approximation} 
\journal{JMR}														

\author[1]{Christopher P. Bidinosti} 
\author[2]{Genevi\`eve Tastevin}
\author[2]{Pierre-Jean Nacher\thanks{nacher@lkb.ens.fr, corresponding author.}}

\affil[1]{Department of Physics, University of Winnipeg, Winnipeg, MB, Canada R3B 2E9.} 
\affil[2]{Laboratoire Kastler Brossel, ENS-Universit\'e PSL, CNRS, Sorbonne Universit\'e, Coll\`ege de France; 24 rue Lhomond, 75005 Paris, France.} 

\twocolumn[         
\begin{@twocolumnfalse}   
\date{}
\maketitle

\vskip -1.0cm 
\hrule
\begin{abstract}
The generation of accurate tip angles is critical for many applications of nuclear magnetic resonance. In low static field, with a linear rather than circular polarized rf field, the rotating-wave approximation may no longer hold and significant deviations from expected trajectories on the Bloch sphere can occur. 
For rectangular rf pulses, the effects depend strongly on the phase of the rf field and can be further compounded by transients at the start and end of the pulse. The desired terminus can be still be achieved, however, through the application of a phase-dependent Bloch-Siegert shift and appropriate consideration of pulse timings. 
For 
suitably shaped rf pulses, the Bloch-Siegert shift is largely phase independent, but its magnitude can vary significantly depending on 
details of the pulse shape as well as the characteristics of the rf coil circuit.
We present numerical simulations and low-field NMR experiments with $^1\rm H$ and $^3\rm He$ that demonstrate several main consequences and accompanying strategies that one should consider when wanting to generate accurate tip angles outside the validity of the rotating-wave approximation and in low static field.

\end{abstract}
\hrule
\vskip 0.5cm
\end{@twocolumnfalse}     
]                         
\saythanks 

\makeatletter
\renewcommand\section{\@startsection {section}{1}{\z@}%
                                   {-3.5ex \@plus -1ex \@minus -.2ex}%
                                   {2.3ex \@plus.2ex}%
                                   {\sffamily\large\bfseries}}
\renewcommand\subsection{\@startsection{subsection}{2}{\z@}%
                                     {-3.25ex\@plus -1ex \@minus -.2ex}%
                                     {1.5ex \@plus .2ex}%
                                     {\sffamily\normalsize\bfseries}}
\renewcommand\subsubsection{\@startsection{subsubsection}{3}{\z@}%
                                     {-3.25ex\@plus -1ex \@minus -.2ex}%
                                     {1.5ex \@plus .2ex}%
                                     {\sffamily\normalsize\bfseries}}
\makeatother

\begin{bibunit}
\section{Introduction}

The vast majority of modern NMR applications operate in a \textit{high-field regime} where the large static field and associated Larmor frequency lead to several simplifying approximations that are often taken as canon. There is, however, a renewed and expanding interest 
in low-field techniques~\cite{Wald2019,Song2019,Budker2019,Marques2019,Blumich2017,Krauss2014} that motivates a much closer look at the physics and practicalities of NMR performed outside the high-field limit. To that end, this paper focuses on experimental conditions where the magnitude of a transverse, linear radio-frequency (rf) field is no longer small compared to that of the static field -- i.e., the breakdown of the 
rotating-wave approximation 
(RWA) -- and where the rf period is no longer small compared to pulse duration. In particular, we explore 
the effects of a strong counter-rotating component of the rf field on the expected tip angle and phase of the transverse magnetization, as well as the dependence of these effects on the specific details of the start and end of the rf pulse.

It is perhaps not surprising that the first authors to consider similar such issues did so many decades ago when the use of more modest field strengths was the rule rather than the exception.  To the best of our knowledge, three central effects studied in this work were first described in seminal papers by the following authors:
Bloch and Siegert (1940), who calculated the shift in resonance field due to the use of non-rotating rf fields for continuous wave (CW) excitation~\cite{Bloch1940};
MacLaughlin (1970), who identified variations in tip angle associated with the counter-rotating term of a linear rf field as well as the start phase of the rf pulse~\cite{MacLaughlin1970}; and Mehring and Waugh (1972), who pointed out the effect of phase transients at the start and end of an rf pulse~\cite{Mehring1972}. 

The potential impact of such effects has been discussed recently in the context of ultra-low field~\cite{Krauss2014,Shim2014} and Earth's field~\cite{Conradi2018} applications of NMR. 
They are of concern also for very low field Ramsey resonance measurements used to search for permanent electric dipole moments of particles such as the neutron~\cite{May1998, Pendlebury2015}. 
Indeed, the breakdown of the RWA is pertinent to any strongly-driven, two-level system~\cite{Benenti2013,Dai2017}, such as a single nitrogen vacancy in diamond~\cite{Fuchs2009,Rao2017} or a qubit~\cite{Zhang2018,castanos2019}.

Practical motivation for this work comes from our interest in potential low-field uses of a new NMR imaging technique known as transmit array spatial encoding (TRASE)~\cite{Sharp2010,Sharp2013,Stockmann2016,Sarty2018,Bohidar2019,Bohidar2020,Sun2020,Nacher2020}. With this method, spatial encoding is achieved through the application of phase gradients
 of the rf field -- rather than magnitude gradients of the static field -- and in general requires as many as two distinct phase-gradient rf coils per encoding 
 direction.\footnote{
The ideal rf field for TRASE has uniform magnitude and linearly varying direction. An example is $\bm B_1(z,t) = B_1\cos(\omega t) \times (\cos(g z)\, \bm{\hat{x}} + \sin(g z)\, \bm{\hat{y}})$ with constant $g$, which one might generate with a spiral coil~\cite{Sharp2010,Nacher2020,Bidinosti2010}. 
 }
  Moreover, traversal through $k$-space with TRASE requires a rapid train of $\pi$-pulses generated by alternate phase-gradient coils. As a result, there is good reason to employ linear rather than rotating rf fields (lest the already large number of rf coils should further double) as well as to understand the generation of short, accurate $\pi$-pulses achieved with such fields. This becomes all the more important, of course, when operating in low static field.

As general interest in low-field NMR grows, the issues related to strong linear rf fields are now receiving a second look. 
Kraus \textit{et al.}~\cite{Krauss2014}, for example, explore the breakdown of the RWA 
 and provide a mathematical formalism to compute the evolution of the magnetization over the course of a rectangular $B_1$ pulse of elliptical or linear rf field. They present simulations of the effects of the counter-rotating component in the low-field regime, but do not consider the impact of the rf phase on these results. 
Mitchell~\textit{et al.}~\cite{Mitchell2014} and Mandal~\textit{et al.}~\cite{Mandal2015}, on the other hand,
review previous work on phase transients and provide further insight into managing the corresponding effects on rf pulse shape and sequence design through simulation and experiment. 
The range of static and rf field magnitudes they explore, however, is such that the breakdown of the RWA is not of fundamental concern. Indeed, the counter-rotating component of the rf field was not included in some simulations~\cite{Mandal2015}. 

In contrast, this work seeks to explore the intricate confluence of all three effects:
Bloch-Siegert shift, phase dependence of the trajectory on the Bloch sphere, and rf
transients. We also explore shaped pulses, and in future work we will consider the effects of the concomitant component of the rf field that lies along the static field direction when phase-gradient coils are used for TRASE imaging~\cite{Nacher2018,Bidinosti2018a}.

This paper is organized as follows. 
In Section~\ref{sec:theory} we introduce the general concepts related to this work -- including the RWA, the Bloch-Siegert shift,
and transients -- in the context of rectangular rf pulses. 
In Section~\ref{sec:simulations} we employ numerical simulations to demonstrate the complex interplay of the many factors that affect the trajectory -- and eventual terminus -- of the magnetization vector on the Bloch sphere. 
In Sections~\ref{sec:experimental} and~\ref{sec:results} we present our experimental methods and results. 
In Section~\ref{Sec6} we summarize our main findings. 
Further experimental details and additional derivations are provided in Supplemental Material (abbreviated as SM)  
at the end of this document.

\section{General Concepts}
\label{sec:theory}

\subsection{The linear rf field, the RWA, and the start phase}
\label{sec:linear_rf}

We begin by providing a suitably general formula for the rf field in the rotating frame during the course of a rectangular rf pulse. 
We assume the laboratory and rotating frames to be coincident at $t=0$, and take the common angular frequency $\omega$ of the rf field and rotating frame to be a positive quantity~\cite{Hoult2000}. For simplicity, we consider the typical case of a linear rf field applied along the $x$-axis of the laboratory frame of the form 
\begin{equation}
\bm B_1(t) = B_1 \cos(\omega t + \phir) \,\bm{\hat{x}} \, ,
\label{B1_lab_cos}
\end{equation}
where $B_1$ is the amplitude, $t$ is absolute time, and $\phir$ is the relative phase of the rf carrier at $t=0$~\cite{Liang1999,Blumich2003,Cowan2010,Corps2011}.
It is not necessary to include $\phir$  in Eq.~\ref{B1_lab_cos}
when contemplating single pulse experiments, and indeed it is often set to zero throughout this work for simplicity. However, in the broader context of composite pulses~\cite{LEVITT198661}, or multi-pulse techniques such as CMPG or TRASE, its inclusion is well justified here. The formulation of Eq.~\ref{B1_lab_cos} 
also affords an opportunity to confront commonly held notions and practices that do not hold outside the RWA. Any further generalization of the rf field~\cite{Hoult2000}, though, is presently unnecessary. 

By simple projection, the components of $\bm B_1(t)$ in clockwise and counter-clockwise rotating frames 
 (as viewed from the $+z$-axis) are 
\begin{eqnarray}
 \bm{\mathcal{B}}_1^\pm(t)&=&\tfrac{1}{2}B_1 \Big( \cos(\phir)+\cos(2\omega t + \phir) \Big) \, \bm{\hat{x}'} \nonumber \\
&\pm&
\tfrac{1}{2}B_1 \Big( -\sin(\phir) +\sin(2\omega t + \phir)  \Big) \,\bm{\hat{y}'} \, ,
\label{B1_rot_cos}
\end{eqnarray}
where the prime distinguishes the unit vectors in the respective rotating frames. 
We denote clockwise (counterclockwise) with a plus (minus) superscript to correspond with the direction of precession 
for a positive (negative) value of the gyromagnetic ratio $\gamma$. On resonance, then, with a static field of magnitude $\omega/|\gamma|$ directed along the $+z$-axis, the dc components of Eq.~\ref{B1_rot_cos} are the familiar RWA fields that drive nutation in the respective cases. For briefness, we explicitly analyze only $\bm{\mathcal{B}}_1^+(t)$ below, relevant for positive-$\gamma$ nuclei.

\begin{figure}[tbh]
\centering
\includegraphics[trim=0 0 0 0, clip=true, width=0.85\columnwidth]{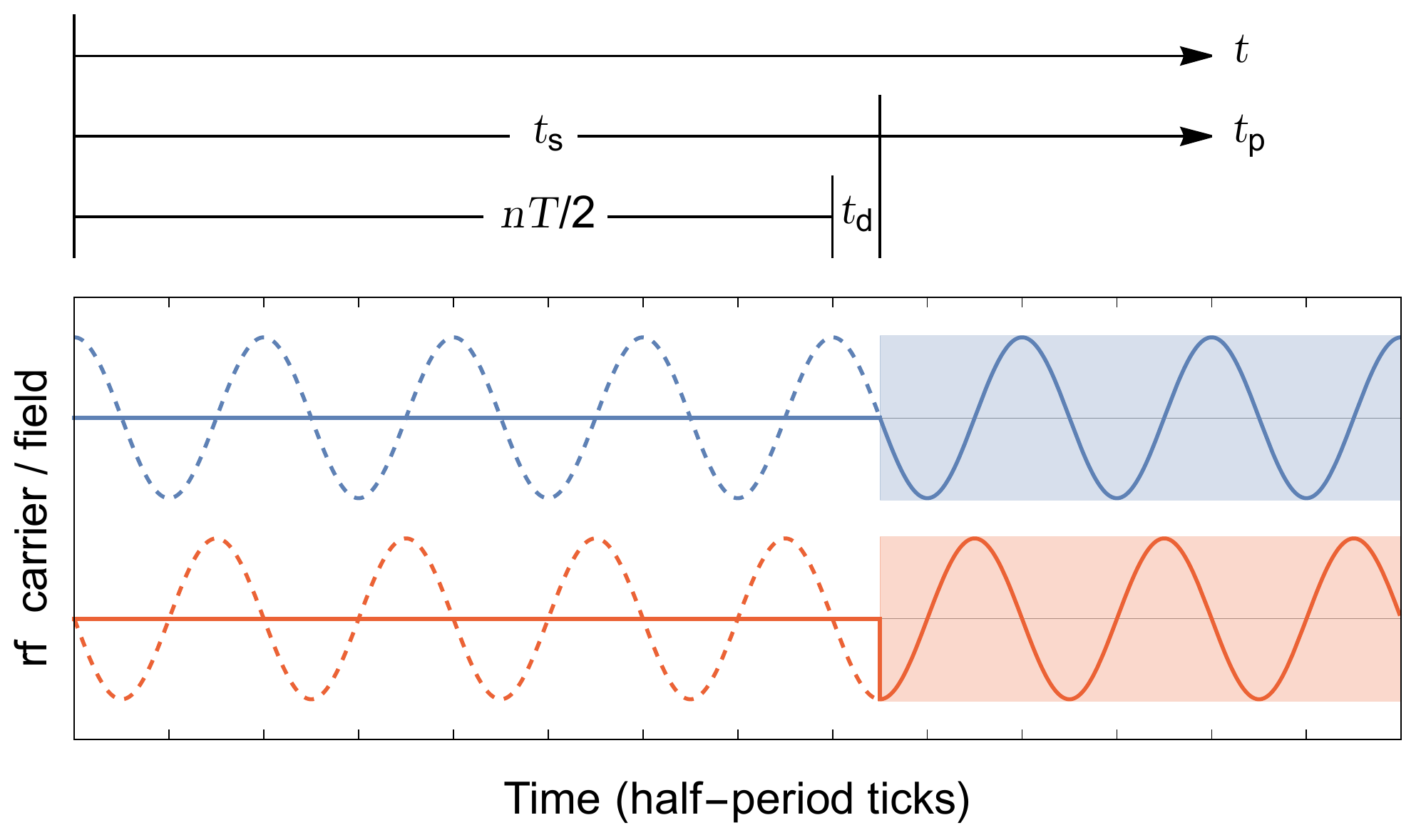} 
\caption{
Timing diagram depicting the rf carrier (dashed lines) and rf field (solid lines) at the start of a rectangular $B_1$ pulse (shaded regions) for $\phir=0^\circ$ (top blue curves) and $90^\circ$ (bottom red curves).  For the scenario sketched here, $n=8$ and $\td=T/4$, giving $\phid=90^\circ$.
}
\label{fig:timings}
\end{figure}

Given the $2\omega$ frequency of the counter-rotating (CR) terms in Eq.~\ref{B1_rot_cos}, the absolute time at the start of a rectangular rf pulse can be written without loss of generality as
$\ts = n\, (T/2) + \td$, where $n$ is an integer, $T$ is the period of the rf field, and $\td$ is a delay 
such that $0<\td< (T/2)$. This is shown in Fig.~\ref{fig:timings}. Letting $\tp$ be the time accrued from the start of the rf pulse and substituting $t=\ts + \tp$ 
into Eq.~\ref{B1_rot_cos} gives the components of $\bm B_1(\tp)$ in the clockwise rotating frame:
\begin{eqnarray}
 \bm{\mathcal{B}}_1^+(\tp) \! \!\! &=& \! \!\!   
 \tfrac{1}{2}B_1 \left( \cos(\phir)+\cos(2\omega \tp + 2\phis - \phir) \right) \,\bm{\hat{x}'} \nonumber \\
\!\!\! &+& \!\! \! 
\tfrac{1}{2}B_1 \left( -\sin(\phir) +\sin(2\omega \tp +2\phis - \phir) \right) \,\bm{\hat{y}'} \, 
\label{B1_rot_td_cos}
\end{eqnarray}
where $\phis = (\phir +\phid)$ is the start phase of the pulse at $t=\ts$ and $\phid=\omega \td$ is a phase delay. 
A key feature to note is that for $\phis=\pm90^\circ$ the rf field is zero at the start of the pulse regardless the value of $\phir$.
This is an important consideration in regard to transients discussed further below.

\begin{figure}[tbh]
\centering
\includegraphics[trim=0 0 0 0, clip=true, width=0.7\columnwidth]{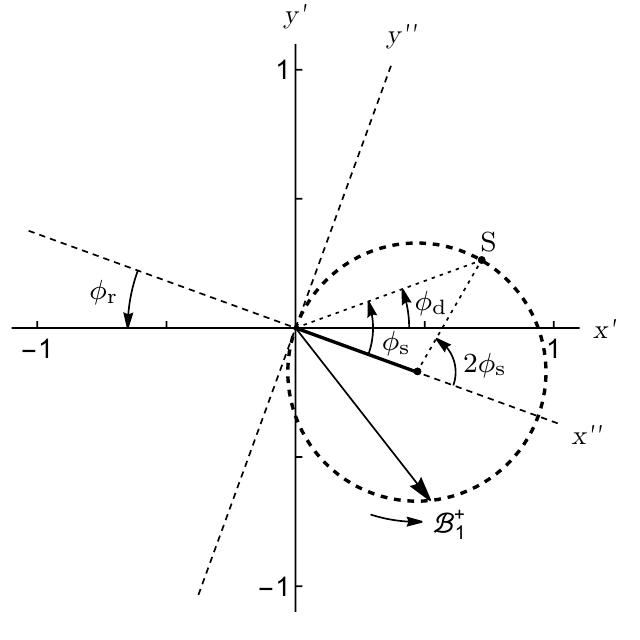} 
\caption{Illustration of a linear rf field of unit magnitude in the rotating frame. The vector $\bm{\mathcal{B}}_1^+(\tp)$ traces a circle (thick dashed line) 
counterclockwise about its dc component (solid line) every half-period.  The latter has a magnitude of 0.5 and lies 
along the positive $x''$-axis.
The start point S on the circle at $\tp=0$ is set by $\phis$ as shown;
it is located at $x''=1$ for $\phis=0^\circ$ and $\pm 180^\circ$, and at the origin for $\phis=\pm 90^\circ$.
For the scenario sketched here, $\phir=20^\circ$, $\phid=20^\circ$, $\phis=40^\circ$, and $\tp =3T/10$.
}
\label{B1plus_vs_time}
\end{figure}

While Eq.~\ref{B1_rot_td_cos} is general, 
and therefore useful when considering multiple pulses made along different axes such as with TRASE,
it is illustrative at this point to make a further change of axis and re-cast it as 
\begin{eqnarray}
\bm{\mathcal{B}}_1^+(\tp) 
&=& \tfrac{1}{2}B_1 \left( 1 + \cos(2\omega \tp +2\phis) \right) \,\bm{\hat{x}''}  \nonumber \\
&+&
\tfrac{1}{2}B_1 \sin(2\omega \tp + 2\phis ) \,\bm{\hat{y}''} \, ,
\label{B1_rot_td_new_cos}
\end{eqnarray}
where $\bm{\hat{x}''}$ 
 gives the direction of the dc component of $\bm{\mathcal{B}}_1^+(\tp)$ in the rotating frame as shown in Fig.~\ref{B1plus_vs_time}. Several important concepts needed for this work become apparent from Eq.~\ref{B1_rot_td_new_cos}. First, as expected, if one invokes the RWA and ignores the CR terms varying as $2\omega \tp$, nutation at resonance occurs in the $x''=0$ plane driven by a 
 constant field $(B_1/2) \,\bm{\hat{x}''}$.\footnote{
If the rf field is not on resonance, precession occurs about an effective field $\bm B_\mathrm{eff}$ that will have a $z$- as well as $x''$-component~\cite{Blumich2003,Cowan2010,Corps2011}.
 }
 If on the other hand the RWA is not valid, and the CR field cannot be ignored, one expects cyclic deviations in the trajectory on the Bloch sphere that occur with a period $T/2$~\cite{Krauss2014,Benenti2013}. Furthermore, since the choice of rotating frame is arbitrary, the particular details of the trajectory depend only on the start phase of the rf pulse and in general can be expected to be unique for each value of $\phis$ modulo $180^\circ$. 
 Indeed as seen from Eq.~\ref{B1_rot_td_new_cos} and Fig.~\ref{B1plus_vs_time}, $\phis$ sets the magnitude and direction of $\bm{\mathcal{B}}_1^+$ at $\tp=0$, and hence the initial influence of the pulse, which subsequently impacts the overall trajectory.
 For example, as noted above, a pulse starting with $\bm{\mathcal{B}}_1^+(0)=0$ can be achieved with $\phis=\pm90^\circ$.

In regard to this last point, we return to Eq.~\ref{B1_rot_td_cos} and point out that one can select any particular nutation axis through the choice of $\phir$ (see Fig.~\ref{B1plus_vs_time}) and subsequently achieve a zero-field start through an appropriate time delay $\td$. For example, to have $\bm{\mathcal{B}}_1^+(0)=0$ with the shortest positive delay in the range $0<\td<(T/2)$, requires the following: 
 \begin{equation}
\td=\frac{\phid}{\omega} =\frac{1}{\omega}
\begin{cases}
(-\pi/2 - \phir) & \text{ if } -\pi \leq \phir \leq -\pi/2 \\
(\pi/2 - \phir) & \text{ if } -\pi/2 < \phir \leq \pi/2 \\
(3\pi/2 - \phir) & \text{ if } \pi/2 < \phir \leq \pi \, . \\
\end{cases} 
\label{eq:td}
\end{equation}
This is, admittedly, a seemingly trivial statement about sinusoidal functions. 
However, its relevance for low-field NMR merits emphasis here.

\subsection{Bloch-Siegert shift}
\label{sec:BlochSiegert}

The work presented here is not related to the
rf amplitude calibration~\cite{Hung2020} and mapping~\cite{Sacolick2010,Sacolick2011,Sturm2012,Clarke2016,Lesch2019} techniques that involve ancillary off-resonance excitation and are generally employed at high field, well within the validity of the RWA.
We are interested, rather, in the breakdown of the RWA for single-frequency excitation and focus on the details of rf-driven nutation under conditions where the effect of the CR component of a linear rf field 
can lead to significant deviation in expected trajectories on the Bloch sphere. 
While our primary concern is with pulsed rf experiments, 
we find it illustrative and useful to frame the discussion around the original CW result of Bloch and Siegert~\cite{Bloch1940}. 

To begin, though, we first recall the basic NMR experiment whereby an rf field of angular frequency $\omega_0 = |\gamma| B_0$ is applied to a sample in a static field $B_0$. Within the validity of the RWA, a rectangular pulse of a linear rf field of amplitude $B_1$ will result in the same trajectory -- and terminus -- of the magnetization vector on the Bloch sphere
 as a rectangular pulse of a circular polarized (CP) rf field of amplitude $B_1/2$ applied for the same duration and with the same phase. 
In either case, the resulting tip angle is $\theta = |\gamma| \, (B_1/2) \, t_\theta$ for a given pulse duration $t_\theta$.
Outside the validity of the RWA, as shown below, the same terminus -- but not trajectory -- can 
 still be achieved with a high degree of accuracy provided that a suitable difference is introduced between the rf frequency and the Larmor frequency. This difference will be referred to here as a Bloch-Siegert shift, in keeping with the definition in the original work~\cite{Bloch1940}.

For simplicity, and for ease of comparison with our typical experimental method, we henceforth limit discussion in this section to the scenario where the rf frequency is held fixed and only the static field is varied. To avoid possible confusion with traditional nomenclature, 
we use $\omega$ to denote the fixed angular frequency of the rf field and its corresponding value in units of magnetic field as $B_\omega = \omega / |\gamma|$. 
Additionally, we denote the general applied static field as simply $B$ and refer to the condition $B=B_\omega$ as resonant (or on-resonance) and $B\ne B_\omega$ as non-resonant (or off-resonance), independent of any other conditions of the experiment and whether the desired terminus is achieved or not.

At this stage we introduce the field shift $\delta B = B_\omega - B$,
which for the original Bloch-Siegert result for CW excitation~\cite{Bloch1940} takes the value
\begin{equation}
\delta B_{\rm cw} = B_\omega \left( \frac{B_1}{4B_\omega} \right)^2  \, .
\label{BS_shift_cw}
\end{equation}
The interpretation of Eq.~\ref{BS_shift_cw} is that one must reduce the static field from $B=B_\omega$ to $B=B_\omega-\delta B_{\rm cw}$ in order to recover the resonant behavior associated with a CP rf field. 
A natural question arises: Does Eq.~\ref{BS_shift_cw} still hold for pulsed rather than CW excitation? In regard to our specific interest in low-field TRASE, one might ask the equivalent question: Does satisfying the condition of Eq.~\ref{BS_shift_cw} allow one to achieve an accurate $\pi$-radian rotation with a pulsed, linear rf field of amplitude $B_1$ in 
the same amount of time as with a resonant CP field of amplitude $B_1/2$? The answer -- shown analytically 
(Appendix~A in SM), 
as well as by simulations (Sec.~\ref{sec:simulations}) and experiments (Sec.~\ref{sec:results}) -- is no. In general the required field shift, which we denote as $\delta B_{\rm p}$, depends on the start phase, amplitude, and shape of the rf pulse. 

To explore this further, we find it convenient to scale pulse durations with respect to the rf period $T=2\pi/\omega$ and in particular introduce the reduced dimensionless parameter 
\begin{equation}
\nu_\pi \equiv \frac{t_\pi}{T} 
\label{eq:nupi}
\end{equation}
for the duration $t_\pi$ of a $\pi$-pulse. Within the RWA, then, the amplitude of a rectangular pulse of linear rf field needed to achieve a $\pi$-radian rotation in $\nu_\pi$ periods is $B_\omega / \nu_\pi$. As such the reciprocal of $\nu_\pi$ also provides an intuitive proxy for rf field magnitude and is conceptually handy when exploring the breakdown of the RWA. 
Furthermore, using this value of the rf amplitude to evaluate the original Bloch-Siegert shift for CW excitation (Eq.~\ref{BS_shift_cw}) leads us to write
\begin{equation}
\delta B_{\rm p} \equiv 
\beta \,   \frac{B_\omega}{16 \, \nu_\pi^2} 
\label{BS_shift_general}
\end{equation}
as the Bloch-Siegert shift for a pulsed excitation.
The prefactor $\beta$, as we shall see,
 depends in a complex way on $\phis$, $\nu_\pi$, and pulse shape.

In general one must rely on numerical simulations, or indeed experiments,
to determine $\beta$ for any particular set of pulse parameters. 
 In some cases, analytic results can be derived (see Appendix~A in SM). 
For example, for a rectangular pulse with $\nu_\pi =n/2$ (i.e. $t_\pi$ equal to an integer multiple $n$ of the rf half-period $T/2$), one finds that the prefactor $\beta$ in Eq.~\ref{BS_shift_general}
is given to lowest order by
\begin{equation}
\beta(\phis) = 1-2\cos 2\phis \, 
\label{beta_rect}
\end{equation}
in the high $\nu_\pi$ (i.e.\ low rf amplitude) limit. To illustrate, some explicit values of $\beta $ in this regime 
 are the following:
 $\mp 1$ for $\phis = 0^\circ$ and $ 45^\circ$, respectively; uniquely zero for $\phis =30^\circ$; 
 and $3$ for $\phis = 90^\circ$. 
 As will be shown in Sec.~\ref{sec:simulations}, higher order effects alter these results and 
 also require an adjustment of the rf amplitude $B_1$ away from the RWA value $B_\omega/ \nu_\pi$ in order to achieve the desired terminus of the magnetization vector.

\subsection{Transients with an untuned rf coil}
\label{sec:phasetrans}

It is important in many NMR applications such as TRASE to generate and repeat rf pulses as quickly as possible. At low field/frequency, a particularly convenient approach is to use an untuned rf coil, which can be treated as a lumped element RL circuit. 
In practice one typically controls the 
driving voltage and not the current in a coil, and it is informative, then, to consider the 
transients that are generated when a rectangular pulse of 
rf voltage is applied to an RL circuit.

It is straightforward to show that if a voltage pulse 
\begin{equation}
V_{\rm rf}(t) = V_{\rm m} \cos(\omega t + \phi_{\rm v})
\label{eq:Vrf}
\end{equation}
is applied to a circuit of resistance $R$ and inductance $L$, starting at $t=0$ with initial conditions $I(0)=0$, the following current is built up in the coil during the voltage pulse:
\begin{equation}
I_{\rm rf}(t) = \frac{V_{\rm m} }{|Z|} \left( \cos(\omega t + \phi_{\rm v} - \phi) - \cos(\phi_{\rm v} - \phi) \, e^{-t/\tau} \right)\, .
\label{eq:currLR}
\end{equation}
Here $Z=\sqrt{R^2+(\omega L)^2} \, e^{i\phi}$, $\phi=\arctan(\omega \tau)$, and $\tau =L/R$. 
Furthermore, when the voltage is subsequently turned off,  the coil current will decay exponentially with time constant $\tau$ from its instantaneous value at that point. 
\begin{figure}[H]
\centering
\includegraphics[trim=0 0 0 0, clip=true, width=0.7\columnwidth]{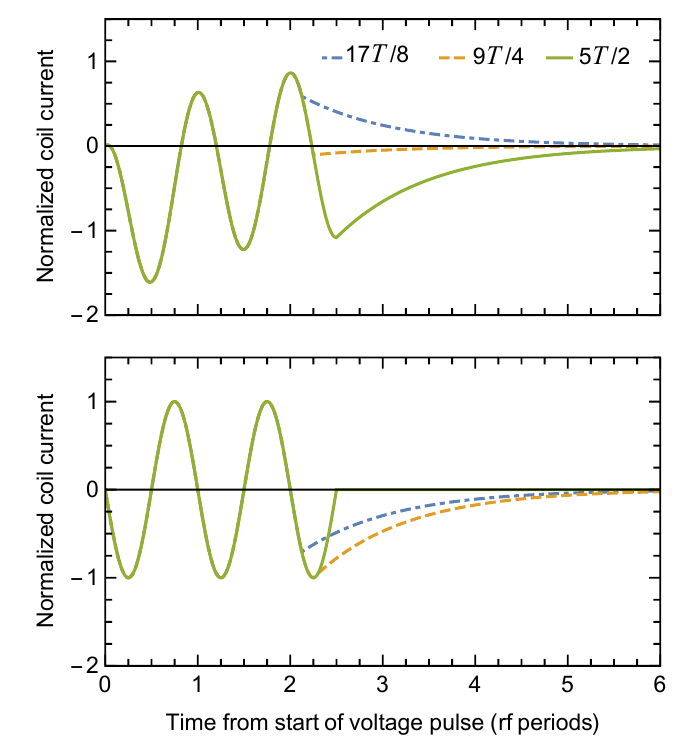} 
\caption{The current generated in a voltage-driven RL circuit with $\tau=T$ for the case $\phis =0^\circ$ (top) and $\phis =90^\circ$ (bottom). 
The 
 rf voltage (not shown) is turned off instantaneously at the times indicated in the legend. }
\label{fig:currents}
\end{figure}
Avoiding transients at both ends of the pulse requires a start phase $\phis = \phi_{\rm v} - \phi = \pm 90^\circ$ and a pulse duration that is an integer multiple of the rf half-period. This is shown in Fig.~\ref{fig:currents}.
There can be many approaches for managing transients in untuned and tuned\footnote{
Note that transients with tuned coils can last much longer, especially for tank circuits with high quality factors.
} 
 rf coils~\cite{Mandal2015,Hoult1979}, of course, but this one is both simple to put into practice (at low frequency) and provides an intuitive link to rectangular pulses that are largely considered 
here.

\section{Numerical Simulations}
\label{sec:simulations}

The goal of this section is to demonstrate through numerical simulations some of the many factors that affect the trajectory and terminus of the magnetization vector on the Bloch sphere. 
We have endeavoured to present a sufficient variety of examples that will highlight the main consequences of the breakdown of the RWA. Parameter space is very large, however, and we stress that one should always employ simulations of their own to guide their particular applications.
 
We ignore the effects of relaxation and diffusion, here, and the resulting Bloch equation 
\begin{equation}
\dot{\bm M}(t) = \gamma \bm M(t) \times (B \,\bm{\hat{z}} + B_1(t)\,\bm{\hat{x}} )
\label{eq:Bloch}
\end{equation}
was solved 
via numerical integration with in-house code (compiled C or Mathematica) to give the magnetization vector $\bm M(t)$ in the laboratory frame. 
For all simulations, we assumed $\gamma>0$. 
For practical reasons, we first chose an rf period $T$ that was typical of our experimental conditions (i.e., tens of micro-seconds) and thereby set $\omega= 2\pi/T$ and $B_\omega=\omega /\gamma$.
``Detuning'' was achieved through a given static field shift $\delta B$ (with the angular frequency $\omega$ of the rf field kept constant), resulting in a net static field $B=B_\omega - \delta B$.
For rectangular pulses, the choice of $\nu_\pi$ sets the pulse duration and the RWA value of the rf amplitude, i.e.\ $B_\omega/\nu_\pi$.
In some studies, we varied $B_1$ in the vicinity of this nominal value through a scale factor of order unity. 
Specific details of shaped pulses are given later.
The remaining input parameters are the relative phase $\phir$ and phase delay $\phid$, which together set the start phase $\phis = \phir+\phid$. Most often we considered the case 
$\phir=0^\circ$ (i.e., the dc component of $ \bm{\mathcal{B}}_1^+(t)$ along $\bm{\hat{x}'} $) with $\phis =\phid =90^\circ$ (i.e., a zero field start).
Following appropriate transformation, trajectories on the Bloch sphere are presented in the 
clockwise rotating frame that is coincident with the lab frame at absolute time $t=0$. 
Ultimately, the results presented here are independent of both the magnitude of the gyromagnetic ratio $|\gamma|$ and the rf period $T$ and are thus completely general.\footnote{Note that trajectories in the counter clockwise rotating frame for $\gamma <0$ would exhibit transverse magnetization components that are simply the complex conjugate~\cite{Levitt1997} of those presented here, i.e.\ $M_{y'} \rightarrow -M_{y'}$.}

Following an introductory example immediately below,
Sections~\ref{sec:IdealRectPulses} and~\ref{sec:TransMag} focus on magnetization reversal and the conditions needed to make an accurate $\pi$-pulse with rectangular $B_1$ pulses,
Section~\ref{sec:shaped} contemplates the same but with shaped 
pulses, while Section~\ref{sec:smalltips} considers small angle tips 
with rectangular pulses. Sections~\ref{sec:shaped} and~\ref{sec:smalltips}
also include studies of the impact of transients in an RL circuit.

\subsection{Consequence of the breakdown of the RWA on rf-driven trajectories}
\label{sec:introSims}

\begin{figure}[tbh]
\centering
\includegraphics[trim=0 0 0 0, clip=true, width=9cm]{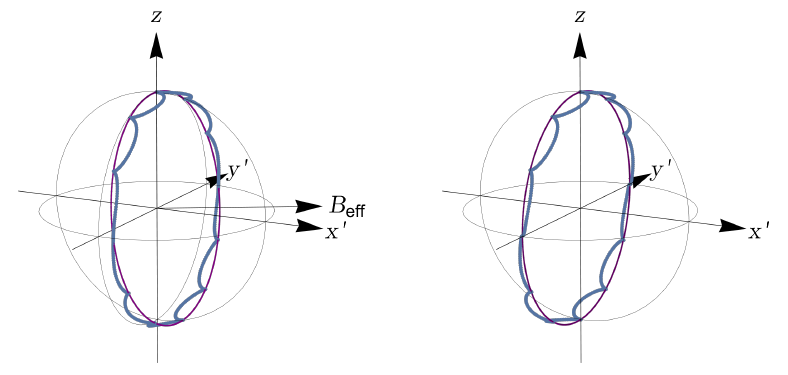} 
\caption{Simulation of trajectories on the Bloch sphere
starting from unit magnetization
driven by a CP rf field (thin purple line) and a linear rf field (thick blue line) with $B_1= (B_\omega/\nu_\pi)$ and $\nu_\pi =3$ as discussed in the text. For the latter, $\phir=0^\circ$ and $\phis =\phid =90^\circ$.
A field shift $\delta B_{\rm p}=3 \, (B_\omega /16 \, \nu_\pi^2)$ of Eqs.~\ref{BS_shift_general} and~\ref{beta_rect}
is applied in the following particular cases.
Left: Nutation via off-resonance CP field ($B=B_\omega+\delta B_{\rm p}$) about an effective field $B_\mathrm{eff}$ (see Appendix~B in SM) 
and on-resonance linear field ($B=B_\omega$). Right: Nutation via on-resonance CP field 
($B=B_\omega$) and off-resonance linear field ($B=B_\omega-\delta B_{\rm p}$). 
}
\label{BS_example}
\end{figure}

An important first example of the breakdown of the RWA is shown in Fig.~\ref{BS_example}
, where time evolution due to rf excitation at constant amplitude is depicted. Here we contrast the trajectory over the course of a $2\pi$-rotation starting from 
unit longitudinal magnetization at the north pole, i.e.\ the point $(0,0,1)$, achieved with either a linear rf field of amplitude $B_1$ or a CP rf field of amplitude $B_1/2$ (which is of course the RWA limit of the former). To accentuate the differences, we choose 
a value of $B_1$ such that the nominal nutation frequency is only one sixth of the resonance frequency (i.e., $B_1= B_\omega/3$ or $\nu_\pi=3$), ensuring significant departure from the RWA.

For the case of the linear rf field, having an amplitude that is no longer negligible compared to the static field, the trajectory on the Bloch sphere becomes notably cycloid-like with cusps occurring every half-period when the instantaneous value of the rf field is zero. Moreover, with $B=B_\omega$, the trajectory does not pass through the south pole, i.e. the point $(0,0,-1)$. It can be made to do so with good accuracy,
 however, through the appropriate shift $\delta B_{\rm p} $ determined from Eqs.~\ref{BS_shift_general} and~\ref{beta_rect}. 
(We shall see later that fine corrections to $\delta B_{\rm p} $ and $B_1$ are needed to exactly pass through the south pole, especially for small $\nu_\pi$ values.)
For the CP rf field, the field shift causes the trajectory to miss the south pole, as expected.
As can also be seen in Fig.~\ref{BS_example}, when
 $\nu_\pi$ is equal to an integer multiple of $1/2$ and $\phis=90^\circ$, the cusps for the trajectories of the linear rf field 
 (on and off resonance) land on the trajectories expected for the CP rf field (off and on resonance). 
 Similar observations were made by Kraus et al.~\cite{Krauss2014}.
 This scenario is of particular interest both conceptually and practically in that it links the zero-current start/stop condition 
 (which avoids transients in an RL circuit) with a terminus that is readily calculated from well-known RWA results.

\subsection{Magnetization reversal from the north pole via rectangular \texorpdfstring{$\bm{B_1}$ $\bm{\pi}$}{B1 Pi}-pulses}
\label{sec:IdealRectPulses}
One can gain further significant insight into magnetization trajectories on the Bloch sphere (and associated deviations from expected RWA results) by simulating rectangular $B_1$ pulses that are switched on/off instantaneously 
\begin{figure}[H]
\centering
\includegraphics[trim=0 0 0 0, clip=true, width=0.79\columnwidth]{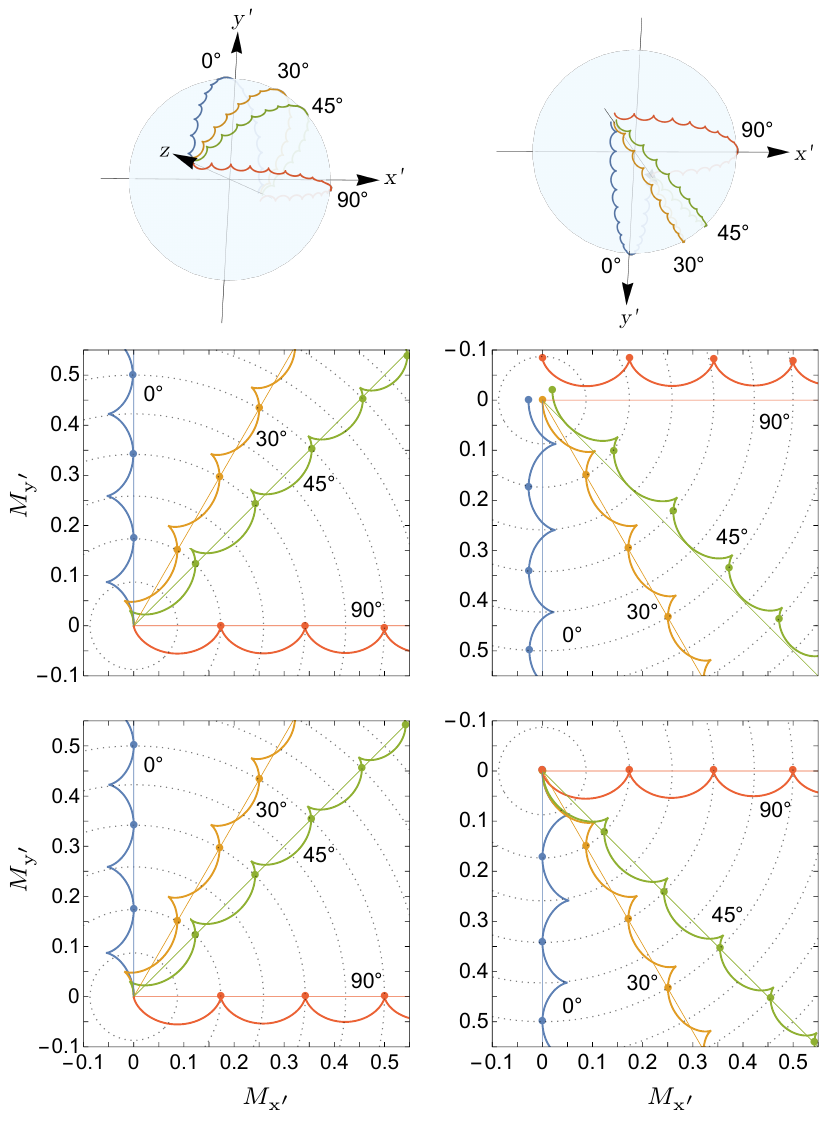} 
\caption{
Simulation of a nominal $\pi$-pulse 
starting from unit longitudinal magnetization at the north pole with $B_1= (B_\omega/\nu_\pi)$, $\nu_\pi =9$ and $\phis=\phir$ (i.e.\ $\phid=0^\circ$) for all cases.  
The start (left) and terminus (right) of the trajectories are shown for $\delta B=0$ (illustrations on Bloch sphere and top graphs) 
and for $\delta B_{\rm p} (\phis)$ of Eqs.~\ref{BS_shift_general} and~\ref{beta_rect}
(bottom graphs). The value of $\phis=\phir$ is indicated next to each trajectory. The components of the transverse magnetization in the rotating frame are plotted in the graphs. The thin straight lines are the corresponding RWA trajectories  for $\delta B=0$ that start 
from the north pole and always end at the south pole. 
The filled circles indicate time steps of $T/2$ over the course of the trajectory. 
The dotted lines indicate constant values of the polar angle $\theta$ in 5$^\circ$ steps.
 }
\label{fig:poles}
\end{figure}
at any point.  Here we focus on conditions that within the validity of the RWA would produce a $\pi$-radian tip angle starting from 
unit longitudinal magnetization at the north pole.

Figure~\ref{fig:poles} shows the distinct trajectory patterns for four different values of the start phase. For clarity, we chose $\td=0$ (i.e., $\phir=\phis$) so that each start phase also leads to a trajectory about a different axis in the rotating frame thereby keeping the curves visibly separated. (This also serves to highlight the function of $\phir$ for choosing a particular nutation axis.) When there is no shift of the static field, only the $\phis=30^\circ$ trajectory has its terminus on the south pole.
However, with the appropriate phase-dependent field shift 
$\delta B_\mathrm{p}(\phis)=\left(1-2 \, \cos 2\phis \right)\times (B_\omega /16 \, \nu_\pi^2)$
of Eqs.~\ref{BS_shift_general} and~\ref{beta_rect}, all trajectories can be made to end at this point. 
Under these conditions, any given trajectory is coincident with the corresponding RWA trajectory (for $\delta B=0$) 
every half-period of the rf field (see Appendix~A in SM). 
However, as is clearly demonstrated here, this only occurs at the location of the cusps (i.e. zero coil current/rf field) for the particular case $\phis=90^\circ$.

\begin{figure}[H]
\centering
\includegraphics[trim=0 0 0 0, clip=true, width=\columnwidth]{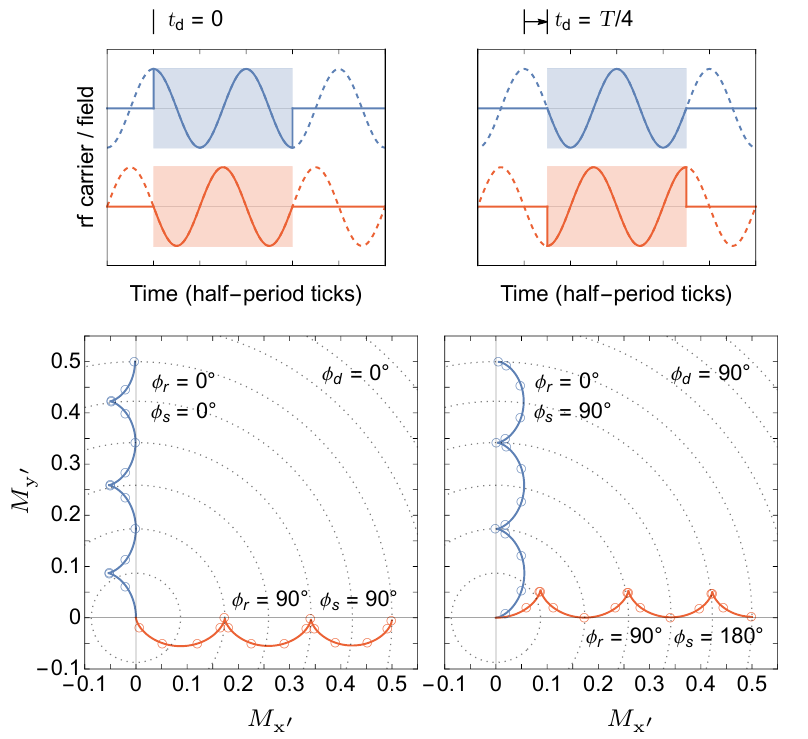} 
\caption{Simulations of rectangular $B_1$ pulses with various combinations of $\phir$ and $\phid$.
 The short pulse duration serves to highlight the impact of $\phis = \phir+\phid$ at the beginning of $\pi$-pulse trajectories. 
Top: The rf carrier (dashed lines) and rf field (solid lines) for a pulse of 
duration $3T/2$ (shaded regions) for $\phir=0^\circ$ (top blue curves) and $90^\circ$ (bottom red curves) for the cases $\td=0$ (left) and $\td=T/4$ (right).
Bottom: Components of the transverse magnetization in the rotating frame for 
simulations starting from unit longitudinal magnetization at the north pole
for each of the above pulses with $B_1= B_\omega/\nu_\pi$ (here $\nu_\pi =9$) and $\delta B=0$.
The values of $\phir$ and $\phis$ are specified next to each trajectory.
The open circles are time steps of $T/10$ along the trajectories; their bunching near the cusps indicates how the nutation stalls at these points when the rf field is zero.
The dotted lines are constant values of the polar angle $\theta$ in 5$^\circ$ steps.
}
\label{fig:startPhase}
\end{figure}

Figure~\ref{fig:startPhase} further emphasizes the effect of the start phase, with trajectories plotted for $\phis=0^\circ$ and $90^\circ$ for two different values of $\phir$ each. 
 Within the validity of the RWA, one expects that any rectangular $B_1$ pulse having the same duration and relative phase $\phir$ will result in an identical trajectory independent of the value of the start phase $\phis$. This is not the case in general, as is shown here. To ensure a specific trajectory pattern (set by $\phis$) about a chosen nutation axis (set by $\phir$) one must employ an appropriate phase delay~$\phid$.

\begin{figure}[htb]
\centering
\includegraphics[trim=0 0 0 0, clip=true, width=\columnwidth]{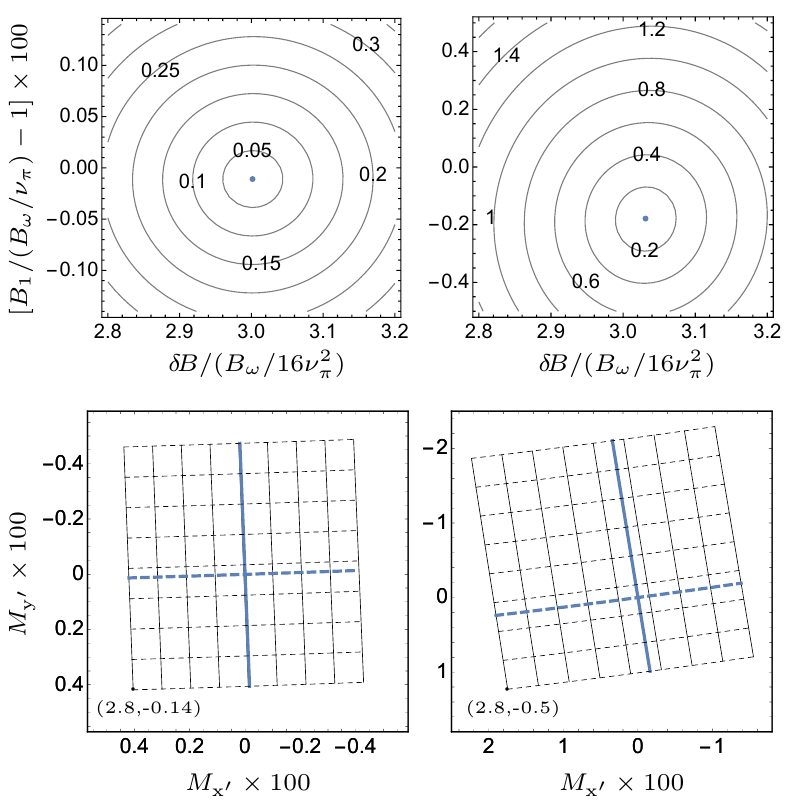} 
\caption{The terminus of a nominal $\pi$-pulse generated with durations given by $\nu_\pi=12$ (left) and $\nu_\pi=3$ (right). For both cases, $\phir=0^\circ$ and $\phis =\phid =90^\circ$.
Top: Contour plots of tip angle error in degrees as a function of rf amplitude and static field shift as discussed in the text.
The blue dot marks the exact $\pi$-pulse.
Bottom: The components of the transverse magnetization at the pulse terminus for the same parameter domain and orientation as above. 
(The parameter coordinates of the bottom left corner of the domain are specified in each graph as a further guide.)
The dashed and solid lines indicate constant values of $B_1$ and $B$, respectively. The bold blue lines indicate the optimal values of the parameter coordinates corresponding to the blue dot above. For $\nu_\pi=12$ these are $B_1 = 0.9999 \, (B_\omega/\nu_\pi)$ and $\beta = 3.002$; for $\nu_\pi=3$ these are $B_1 = 0.9982 \, (B_\omega/\nu_\pi)$ and $\beta = 3.032$.
}
\label{fig:fine_tune}
\end{figure}

In Fig.~\ref{fig:fine_tune}, we present a closer look at the conditions required to make an exact $\pi$-pulse starting from 
unit longitudinal magnetization at the north pole.
In particular, we explore the resulting terminus for a pulse of fixed duration $\nu_\pi T$ for small variations about both the RWA rf amplitude $B_\omega/\nu_\pi$ and the lowest-order static field shift given by Eqs.~\ref{BS_shift_general} and~\ref{beta_rect}. 
It is found that for a fixed pulse duration, the exact $\pi$-pulse occurs for a unique pair of these parameters. For modest pulse strengths (e.g., $\nu_\pi =12$), the use of the nominal values (i.e., $B_1= B_\omega/\nu_\pi$ and $\beta = 3$ for $\phis=90^\circ$ presented here) is very likely more than sufficient to achieve the desired accuracy of the terminus. However, for stronger pulse strengths (e.g., $\nu_\pi =3$), one may wish to make adjustments. 

From a practical point of view, one can also see from Fig.~\ref{fig:fine_tune} that (i) varying the rf amplitude largely results in a change in terminus along the direction corresponding to the RWA trajectory (i.e., $\bm{\hat{y}'}$ here) and (ii) varying the static field shift
largely results in a change in terminus along the direction of the dc component of $\bm{\mathcal{B}}_1^+$ (i.e., $\bm{\hat{x}'}$ here). This provides 
a means for finding the conditions that give an exact $\pi$-pulse. 

\begin{figure}[bth]
\centering
\includegraphics[trim=0 0 0 0, clip=true, width=\columnwidth]{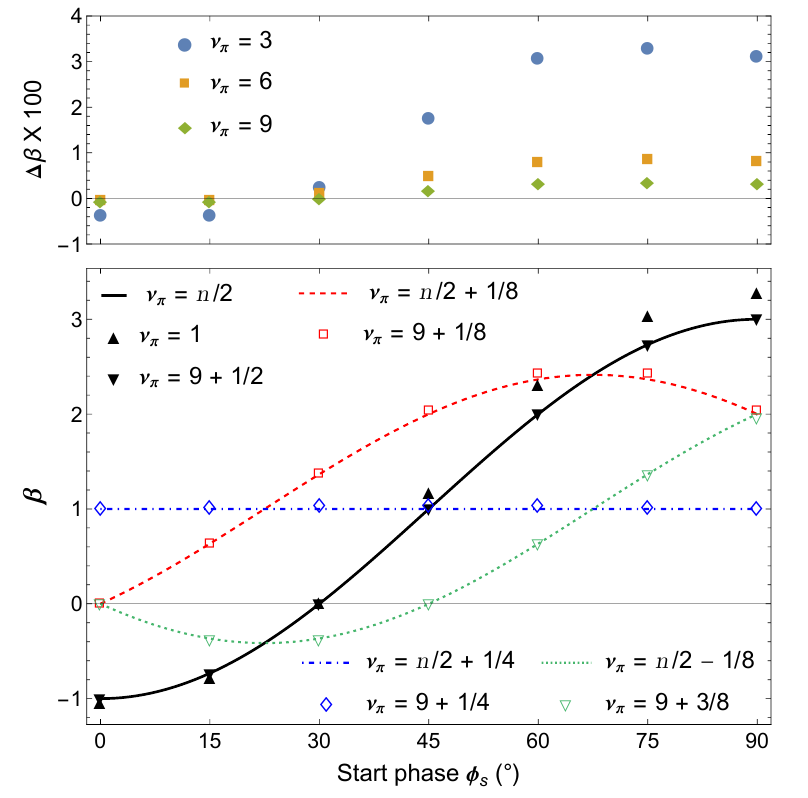} 
\caption{Bottom: Lowest-order 
analytic results of Eq.~A.16 in SM 
(curves) versus values for $\beta(\phis,\nu_\pi)$ extracted from simulations that achieve an accurate $\pi$-pulse (symbols).
Numerical results for $\nu_\pi=1$ versus $9+1/2$ highlight the magnitude of higher order effects for the $\nu_\pi = n/2$ case. 
Top: The difference between the numerical and lowest-order analytic result for $\nu_\pi=3$, 6, and 9.
}
\label{fig:prefactor}
\end{figure}

We next explore the conditions needed to achieve an accurate $\pi$-pulse for a wider range of start phases and pulse durations. 
As above, we vary both the rf amplitude and the static field shift to achieve the exact $\pi$-pulse.
We focus our discussion here on results for $\delta B_\mathrm{p}$, which are shown in terms of the prefactor $\beta(\phis,\nu_\pi)$ in Fig.~\ref{fig:prefactor}. Two important features are seen here. First, deviations from Eq.~\ref{beta_rect} for the case where $\nu_\pi = n/2$ indicate the magnitude of effects beyond first order in $1/\nu_\pi$ (see Appendix~A in SM). 
For $\nu_\pi =1$, for example, the optimal value of $\beta$ can be as much as 10\% greater than the lowest-order analytic result, depending on the start phase $\phis$. Differences fall below the $ 1\%$-level by $\nu_\pi = 6$, however.  Second, as noted in Appendix~A in SM), 
when $\nu_\pi \neq n/2$, the form of $\beta(\phis$) can vary significantly from that given by Eq.~\ref{beta_rect}. For example, when $\nu_\pi = n/2 + 1/4$, the required static field shift to achieve an accurate $\pi$-pulse is, to lowest order, just $B_\omega /(16 \, \nu_\pi^2)$, i.e. $\beta=1$, independent of $\phis$.

\begin{figure}[H]
\centering
\includegraphics[trim=0 0 0 0, clip=true, width=\columnwidth]{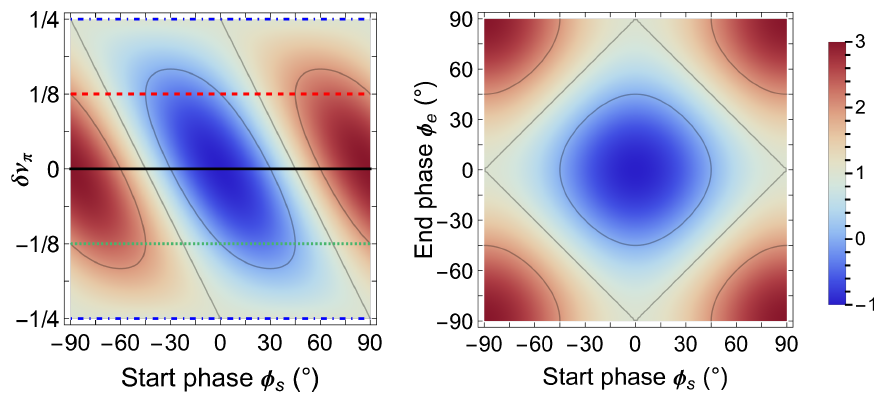} 
\caption{Density plots of the lowest-order analytic solution for $\beta$ from Eq.~A.15 in SM 
as a function of start phase and pulse duration difference (left) or end phase (right). Contour lines (gray) are the values $\beta=0, 1, 2$. The analytic results of Fig.~\ref{fig:prefactor} are shown on the left with the corresponding line color and style. }
\label{beta_global}
\end{figure}

In Fig.~\ref{beta_global}, we present density plots of $\beta$ over the entire range of start and end phases of a rectangular $B_1$ pulse for the lowest-order analytic solution from Eq.~A.15 in SM, 
which was also verified with simulations. The phase at the end of the pulse can be written as $\phie = \phis + 2 \pi \, (n/2 +\delta \nu_\pi)$, where $\delta \nu_\pi$ is the difference in pulse duration from the case $\nu_\pi = n/2$. This provides an intuitive link to the preceding discussion of this section, in particular the results shown in Fig.~\ref{fig:prefactor}. Of notable interest in Fig.~\ref{beta_global} is the contour $\beta =0$, set by the constraint $\cos^2 \!\phis + \cos^2\! \phie = 3/2$. Any rectangular pulse with start and end phases satisfying this condition will, to lowest order, generate a $\pi$-pulse without any static field shift required.

\subsection{Rectangular \texorpdfstring{$\bm{B_1}$ $\bm{\pi}$}{B1 Pi}-pulses applied to transverse magnetization}
\label{sec:TransMag}

Many applications, such as CPMG experiments~\cite{Blumich2003,Cowan2010,Corps2011} or TRASE~\cite{Sharp2010,Sharp2013,Stockmann2016,Sarty2018,Bohidar2019,Nacher2020}, require $\pi$-pulses applied to the magnetization in the transverse plane. Simulations of such a case are presented here without and with a static field shift for $\nu_\pi =9$. The longitudinal magnetization $M_z$ prior to the pulse is zero. The initial phase of the transverse magnetization $M_{\perp'}$, measured relative to the direction of the dc component of $ \bm{\mathcal{B}}_{\rm rf}^+$ (i.e., $\bm{\hat{x}'}$ here), varies in $15^\circ$ steps from $0^\circ$ to $180^\circ$.

\begin{figure}[h]
\centering
\includegraphics[trim=0 0 0 0, clip=true, width=\columnwidth]{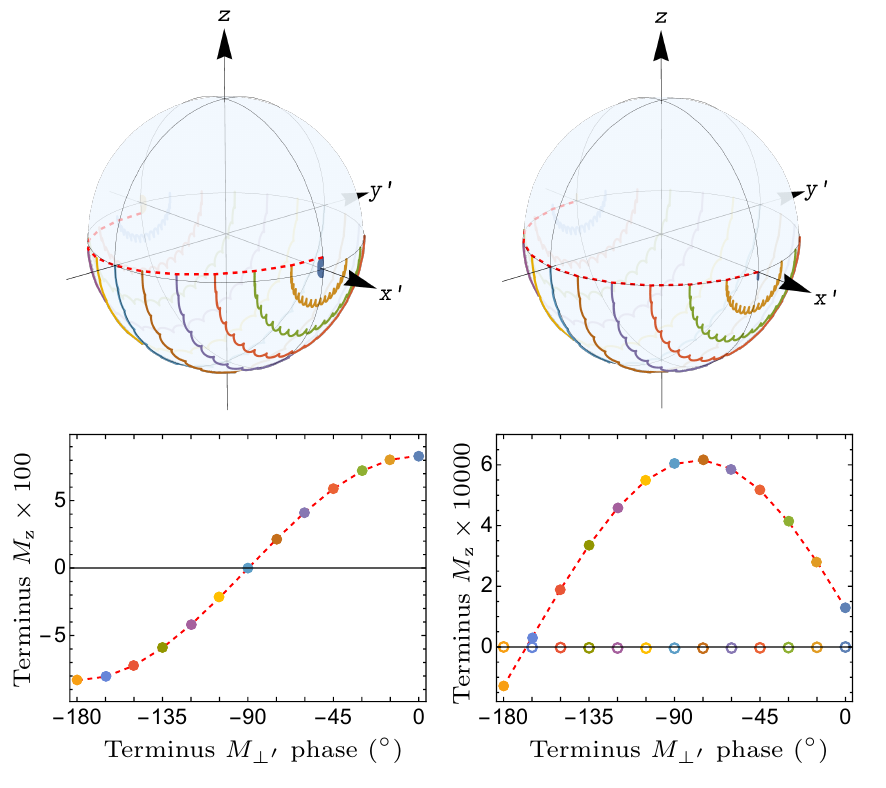} 
\caption{Simulations of a $\pi$-pulse with $\nu_\pi =9$ and $B_1=B_\omega / \nu_\pi$ starting from unit magnetization in the transverse plane. Results are shown for $\delta B=0$ (left) and $\delta B_{\rm p}(\phis)$
of Eqs.~\ref{BS_shift_general} and~\ref{beta_rect} (right). For both cases, $\phir=0^\circ$ and $\phis =\phid =90^\circ.$ 
Top: Trajectories on the Bloch sphere. The initial values of the phase of the transverse magnetization $M_{\perp'}$ are in steps of $15^\circ$  from $0^\circ$ to $180^\circ$, 
counterclockwise from the positive $x'$-axis. The dashed red line connects the terminus of each of the 13 simulations and serves as a visual guide.
Bottom: Terminus values of the longitudinal magnetization $M_z$ versus terminus phase of $M_{\perp'}.$ The data points are connected by the same guide line from above. 
The open symbols at right are for the optimal values of $B_1$ and $\beta$ given in the text.
}
\label{fig:EastWest}
\end{figure}

Figure~\ref{fig:EastWest} shows the magnetization trajectories over the course of the $\pi$-pulse, as well as the terminus values of $M_z$. With no static field shift, the amount of over/undershoot following the pulse (as determined by $M_z$) depends on the initial phase of $M_{\perp'}$ and can be as much as 8\% for this particular example. In contrast, the phase of $M_{\perp'}$ at the terminus varies by only $0.05^\circ$ or less from the expected RWA result, which in this case is simply the negative value of the initial phase. 
When the lowest-order static field shift (i.e. $\beta =3$ here) is applied, the amount of over/undershoot is reduced by two orders of magnitude and phase differences become negligible. If instead one uses the optimal values for the rf amplitude ($B_1= 0.9998 \, B_\omega/\nu_\pi$) and static field shift ($\beta = 3.0035$) found from simulations as per the previous section, the terminus values of $M_z$ are at the ppm level or less.

\subsection{Magnetization reversal from the north pole via shaped \texorpdfstring{$\bm{\pi}$}{Pi}-pulses}
\label{sec:shaped}

\begin{figure}[tbh]
\centering
\includegraphics[trim=0 0 0 0, clip=true, width=0.9\columnwidth]{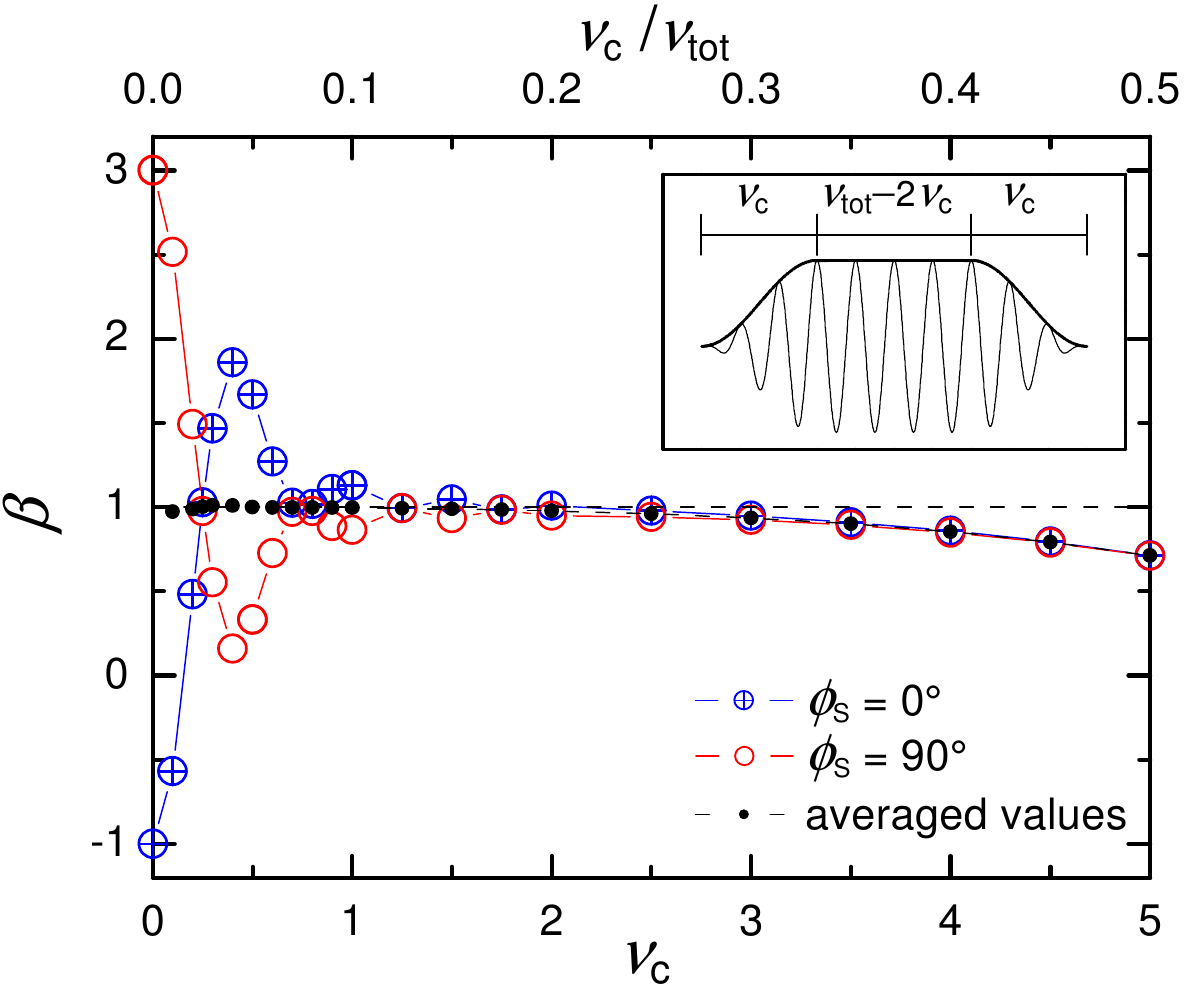} 
\caption{Simulation data
 for 10-period-long shaped $B_1$ $\pi$-pulses.
 Inset: example of a pulse
of amplitude $B_1(t)$ of Eq.~\ref{CPC_piecewise} with $\nu_{\mathrm{c}}=3$ and $\nu_{\mathrm{p}}=4$. Main plot: extremal and averaged
Bloch-Siegert shift prefactors for start phases $\phis=0^\circ$ and~$90^\circ$
are plotted versus $\nu_{\mathrm{c}}$ (lower scale) and $\nu_{\mathrm{c}}%
/\nu_{\mathrm{tot}}$ (upper scale). The computed shifts are scaled to the
CW values associated with the maximum amplitudes of the applied pulses
(Eq.~\ref{BSCW-max}) to give $\beta$.}
\label{Fig-CPC-Volts}
\end{figure}

In order to alleviate the consequences of an instantaneous pulse start/stop, 
we now consider a slowly varying time-dependent amplitude $B_1(t)$
in lieu of the constant value in Eq.~\ref{B1_lab_cos}.
 We choose a simple, cosine-like transition function lasting $\nu_{\mathrm{c}}$ rf periods on either side of a plateau region of $\nu_{\mathrm{p}}$ periods. The overall rf envelope is defined piecewise as
\begin{equation}
B_1(\tp) = \frac{B_{\mathrm{1max}}}{2}
\begin{cases}
  [1-f(\tp)] , 		& \text{for } 0 \leq \tp < t_{\mathrm{p1}} \\
  2, 						& \text{for } t_{\mathrm{p1}} \leq \tp <  t_{\mathrm{p2}}  \\
   [1+f(\tp- t_{\mathrm{p2}} )], 	& \text{for } t_{\mathrm{p2}}   \leq  \tp \leq \nu_{\mathrm{tot}}T
 \end{cases}
\label{CPC_piecewise}
\end{equation}
where $f(t) = \cos \left( \pi t/ \nu_{\mathrm{c}} T \right)$, the transition times are $t_{\mathrm{p1}} = \nu_{\mathrm{c}} T$ and $t_{\mathrm{p2}} = (\nu_{\mathrm{tot}} - \nu_{\mathrm{c}} ) T $, and $\nu_{\mathrm{tot}}= 2\nu_{\mathrm{c}} +\nu_{\mathrm{p}}$ is the total number of rf periods (not necessarily an integer) comprising the pulse. Equation~\ref{CPC_piecewise} yields a smooth rise to, and decay from, a central plateau of constant
amplitude $B_{\mathrm{1max}}$ with null derivatives at its boundaries. An example is shown in the inset of Fig.~\ref{Fig-CPC-Volts}
for $\nu_{\mathrm{tot}}=10.$

 We focus here on the conditions needed to generate an accurate $\pi$-pulse. 
Trajectories on the Bloch sphere similar to those displayed in
Fig.~\ref{fig:poles} are obtained, however with smaller deviations from RWA
trajectories near the start and terminus, where the rf field has a weaker
amplitude. 
As is done for a rectangular pulse, we define a prefactor $\beta$ that scales the static field shift 
$\delta B_\mathrm{p}$ (needed to perform an exact $\pi$-pulse) to the CW value of the Bloch-Siegert shift 
evaluated for a relevant rf amplitude. For reasons which will appear further in this section, we
choose to characterize shaped pulses by the amplitude
$B_{\mathrm{1max}}$ of their plateau. The nominal $\pi$-radian rotation
condition for a rectangular pulse, i.e.\ $\nu_{\pi}B_{1}=B_{\omega}$, is thus replaced
with $\nu_{\mathrm{tot}}\left\langle B_1(\tp)\right\rangle
=B_{\omega}$ for a shaped pulse, where the brackets denote time-average.
Noting that
\begin{equation}
\left\langle B_1(\tp)\right\rangle =\left( 1-\nu_{\mathrm{c}
}/\nu_{\mathrm{tot}}\right) B_{\mathrm{1max}},\label{avtomax}
\end{equation}
substitution into Eq.~\ref{BS_shift_cw} leads us to write
\begin{equation}
\delta B_\mathrm{p} = \beta \left( 
\frac
{B_{\omega}}{16\left( \nu_{\mathrm{tot}}-\nu_{\mathrm{c}}\right) ^{2}} \right) \, ,
\label{BSCW-max}
\end{equation}
for a shaped pulse of Eq.~\ref{CPC_piecewise},
where $( \nu_{\mathrm{tot}}-\nu_{\mathrm{c}} ) $ is the width in periods of the rf field half-maximum. 

Using this expression, $\phis$-dependent $\beta$ factors are derived from computed Bloch-Siegert shifts. 
They are plotted in
Fig.~\ref{Fig-CPC-Volts} for $\nu_{\mathrm{tot}}=10$ for the two bounding cases $\phis=0^\circ$ and~$90^\circ$. The large dependence on
the start phase obtained for rectangular pulses is rapidly reduced for shaped
pulses with increasing transient durations. Following an inverted sign for
$\nu_{\mathrm{c}}=0.4,$ the span of $\beta$ values is for instance divided by
14 for $\nu_{\mathrm{c}}=1$ and by 56 for $\nu_{\mathrm{c}}=2.$

Additional simulations performed for longer shaped pulses
($\nu_{\mathrm{tot}}=15$ and 20, not displayed) show that this damped
oscillatory behavior of $\beta$ essentially depends on the value of
$\nu_{\mathrm{c}},$ i.e. on the absolute steepness of the amplitude
transients. On the contrary, the average values (and actual values above
$\nu_{\mathrm{c}}\approx2$) essentially depend on the aspect ratio
$\nu_{\mathrm{c}}/\nu_{\mathrm{tot}}$ (the upper scale in
Fig.~\ref{Fig-CPC-Volts}). This plot can therefore be used to evaluate
expected shifts for shaped pulses of any duration.

As discussed in Sec.~\ref{sec:phasetrans}, when a voltage pulse is applied
to an untuned rf coil with a characteristic time constant $\tau,$ the resulting current and
therefore rf field may differ significantly from the $\tau=0$
response assumed so far in this section. For the purpose of simulations, currents can be computed 
for a given voltage pulse via convolution with the impulse response of the RL circuit:
\begin{equation}
I_{\mathrm{rf}}(\tp) = \int_{0}^{\tp} \frac{e^{-t/\tau}}{\tau} \,
\frac{V_{\mathrm{rf}}(\tp-t)}{R} dt \, , \label{Icpc}
\end{equation}
where $\tau=L/R$ is the time constant.
Here we assume an rf voltage $V_{\mathrm{rf}}(\tp)$ with an amplitude that is defined piecewise as in 
Eq.~\ref{CPC_piecewise} and multiplied by a carrier part $\cos\left(\omega\tp+ \phi_{\rm v}\right)$. 
The resulting current pulse is calculated over the range $0 \leq \tp \leq \nu_{\mathrm{tot}} T + 5 \tau$, allowing sufficient time to explore the effects of transients.

For small $\nu_{\mathrm{c}}$ and large $\tau/T$
(e.g.\ 1 and 5, respectively), large slowly damped DC components appear,
similar to those in Fig.~\ref{fig:currents}. There is no clearly defined total
duration of the current pulse any more, but numerical demodulation with
suitable filtering reveals that the ac component has a magnitude whose shape is very similar
to that of the amplitude of the driving voltage. 
It is delayed and artefacts appear for small amplitudes, but the half-maximum
width $(\nu_{\mathrm{tot}}-\nu_{\mathrm{c}})$ is a very robust characteristic of all pulse shapes. This is the
reason of the choice of Eq.~\ref{BSCW-max} to scale computed Bloch-Siegert
shifts as well as experimental data in Sec.~\ref{sec:results}.

\begin{figure}[tbh]
\centering
\includegraphics[trim=0 0 0 0, clip=true, width=0.75\columnwidth]{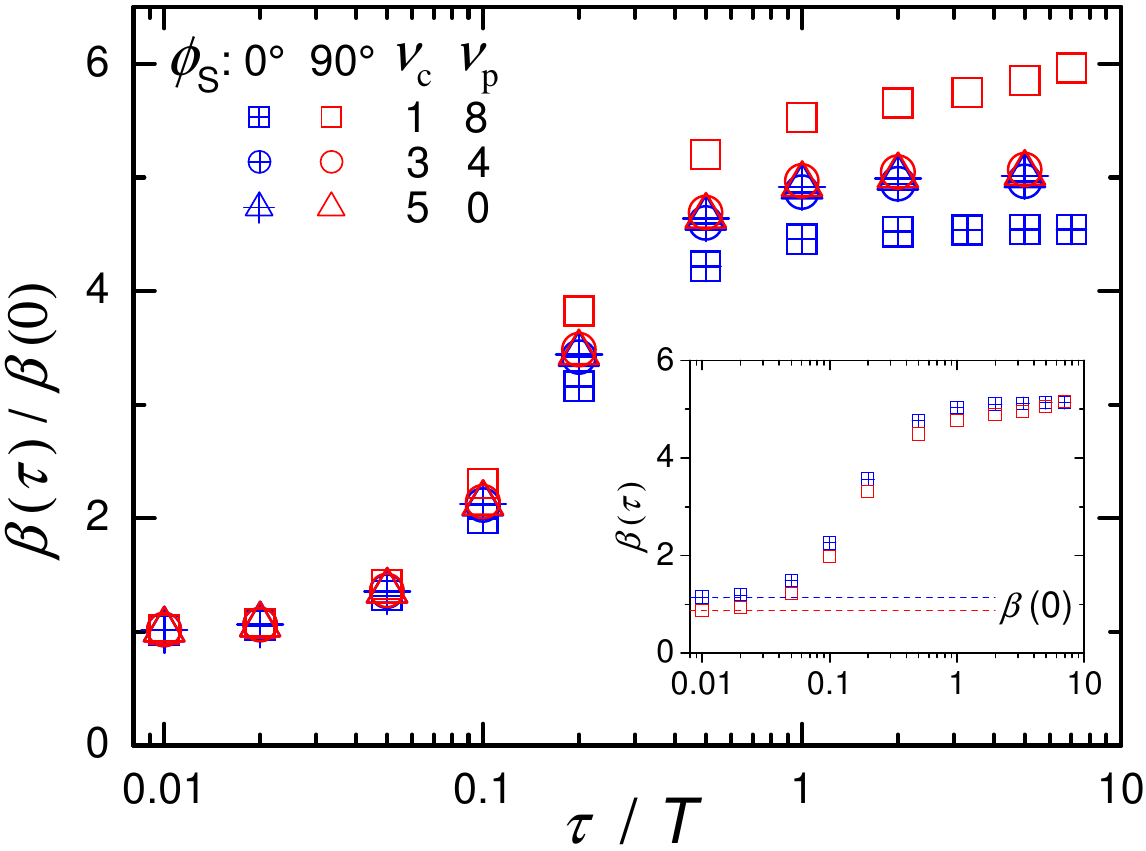} 
\caption{Simulation data for currents resulting from 10-period-long shaped pulses (Eq.~\ref{Icpc}) plotted versus the reduced RL circuit time constant. Different voltage pulse shapes and start phases are considered (see the legend). At
fixed $\nu_{\mathrm{c}}$ (i.e., fixed $\nu_{\mathrm{p}} =10 - 2 \nu_{\mathrm{c}}$) and $\phis$, the values of $\beta(\tau)$ are scaled
to the $\tau=0$ value found for the corresponding shaped $B_1$ pulses of Fig.~\ref{Fig-CPC-Volts}. Inset:
unscaled data for $\nu_{\mathrm{c}}=1.$}
\label{Fig-CPC-LR}
\end{figure}

Series of simulations were performed using such computed currents, and values of the prefactor $\beta$ needed to generate accurate $\pi$-pulses were determined. Select results are shown in Fig.~\ref{Fig-CPC-LR} as a function of $\tau$ for shaped pulses all with $\nu_{\mathrm{tot}}=10$ but composed of different proportions of $\nu_{\mathrm{c}}$ and $\nu_{\mathrm{p}}$. The inset displays examples of $\beta(\tau)$ values for which the small effect of $\phis$ at $\tau=0$ (see $\nu_{\mathrm{c}}=1$ in Fig.~\ref{Fig-CPC-Volts}) is found to result in a near constant span of $\beta$ for non-zero values of $\tau$. 
For slower rise/fall times of the voltage pulse (i.e. $\nu_{\mathrm{c}} > 1$), $\beta$ is found to be largely independent of $\phis$ for any value of the RL circuit time constant, not just for $\tau=0$ as presented in Fig.~\ref{Fig-CPC-Volts}.

All results are more clearly discussed using ratios of $\beta(\tau)$ to $\beta(\tau \! = \! 0)$
that are compiled in the main plot of Fig.~\ref{Fig-CPC-LR} for each pulse shape and bounding value of $\phis$. 
An enhancement of $\beta$-values
is observed for all pulse shapes, with an onset occurring at 
$\tau/T\approx0.05$ and a saturation above $\tau/T\approx0.5.$ For $\nu_{\mathrm{c}}=1$ 
the increasing difference of scaled data obtained for different $\phis$
corresponds to the near constant difference of the unscaled values, but the
averages over $\phis$ are in line with the $\phis$-independent scaled data of
the other voltage pulse shapes.

\subsection{Shortened rectangular pulses for small tip angles -- without and with RL circuit transients}
\label{sec:smalltips}

We now examine the effects of the breakdown of the RWA on small tip angle pulses, i.e.\ $\theta \leq \pi/2$. Such pulses are useful for $T_1$ relaxation measurements~\cite{Cowan2010,Corps2011}, for example, while a $\pi/2$-pulse is used to initiate several imaging sequences and is especially important for TRASE. Within the validity of the RWA, and at high field/frequency where the impact of transients is typically negligible, one can simply shorten by an appropriate factor a previously calibrated $\pi$-pulse, say, to achieve any desired tip angle less than this. Outside the RWA, such a strategy fails in general. We demonstrate this first with rectangular $B_1$ pulses (i.e.\ instantaneous rise and fall of the rf field). We then include the effect of transients by performing simulations with the time dependent $B_1$ fields driven by rectangular pulses of 
rf voltage applied to RL circuits (see Section~\ref{sec:phasetrans}).
Only the case of starting from unit longitudinal magnetization is treated here.

\begin{figure}[htb]
\centering
\includegraphics[trim=0 20 0 0, clip=true, width=0.77\columnwidth]{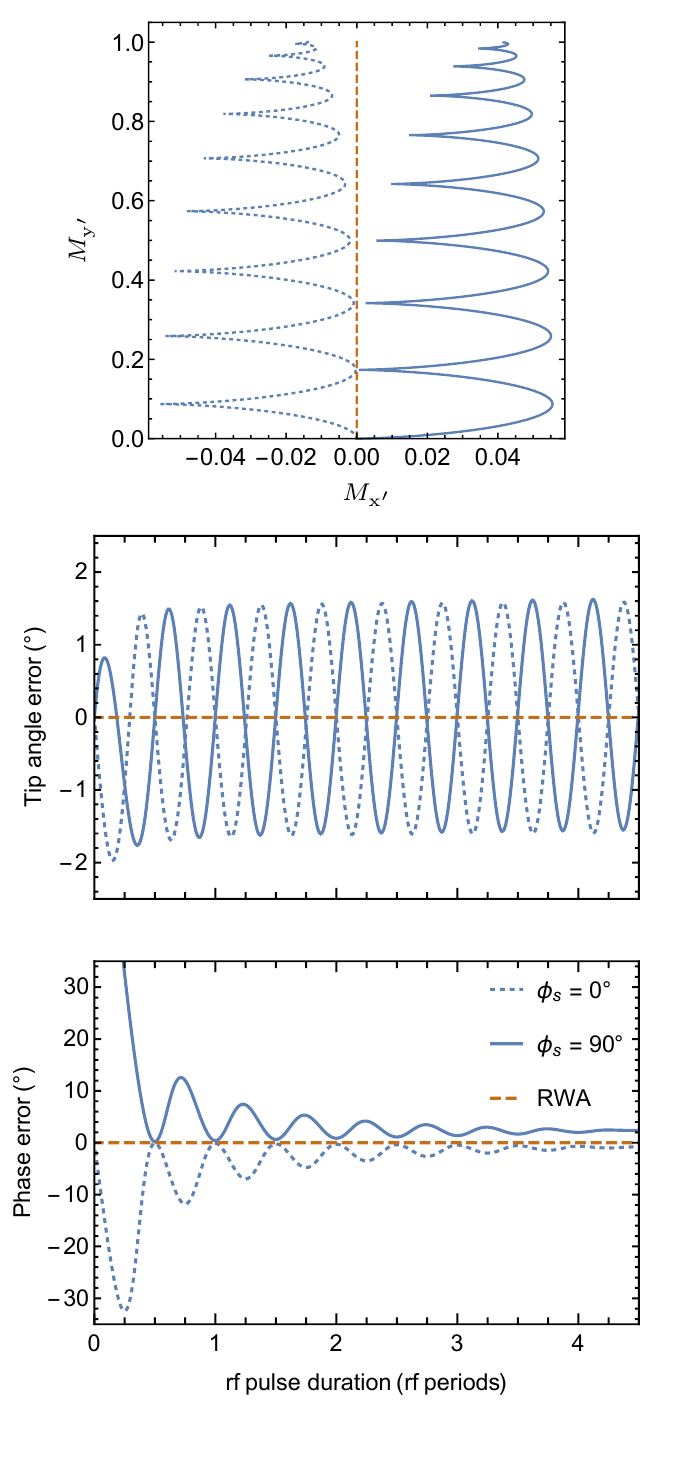} 
\caption{Simulations of a nominal $\pi/2$-pulse starting from unit longitudinal magnetization at the north pole with 
$B_1= B_\omega/\nu_\pi$ (here $\nu_\pi =9$), $\delta B=0$, $\phir =0^\circ$, and $\phis =\phid$ as given in the legend. 
Components of the transverse magnetization in the rotating frame over the course of the trajectory (top) along with tip angle error (middle) and 
$M_{\perp'}$ phase error  (bottom) with respect to the RWA result. The legend refers to all graphs. 
}
\label{small_angle_traj}
\end{figure}
The top graph of Fig.~\ref{small_angle_traj} shows the trajectory over the course of a nominal $\pi/2$-pulse for two different start phases $\phis = 0^\circ$ and $90^\circ$, corresponding to a maximal and zero rf amplitude at the start of the pulse, respectively.  Given that we are considering rectangular $B_1$ pulses here, one can also take any point along the trajectory as the terminus of a pulse of shorter duration. A comparison to the RWA trajectory at the equivalent point in time gives the tip angle and phase errors associated with all nominal pulses over the entire range. These are shown in the bottom two graphs of Fig.~\ref{small_angle_traj}. 

For the example presented here, the tip angle error oscillates between $\sim \pm 1.5^\circ$ over the full pulse duration. 
As a result, small tip angle pulses will tend to suffer the largest relative errors. They also tend to suffer larger phase errors in the transverse magnetization $M_{\perp'}$. For every half-period duration of the rf pulse, the tip angle error is zero and occurs at the same time as a local minimum in the phase error. When the appropriate field shift is applied, the latter will also be zero. 

With or without a static field shift, it is advantageous, then, to choose an integer number of half-periods of the rf field to achieve the same, easily-computed tip angle $\theta$ expected from the RWA. 
For such a pulse duration, i.e. $t_\theta=n\, (T/2)$ with $n$ an integer, the corresponding nominal rf amplitude $B_\omega/\nu_\pi$ is given by $\nu_\pi=n \pi/(2 \theta)$, with $\theta$ in radians.  There is no restriction on the value of the real number $\nu_\pi$ here. However, integer values of $\nu_\pi$ that are multiples of 2 or 3, say, do lead to convenient values of tip angle expressed in degrees (e.g.\ $\theta=10^\circ$, achieved with $n=1$ and $\nu_\pi=9$). These would also have the corresponding $\beta$ factors given by Eq.~\ref{beta_rect} (to lowest order in $1/\nu_\pi$).

\begin{figure}[tbh]
\centering
\includegraphics[trim=0 0 0 0, clip=true, width=0.95\columnwidth]{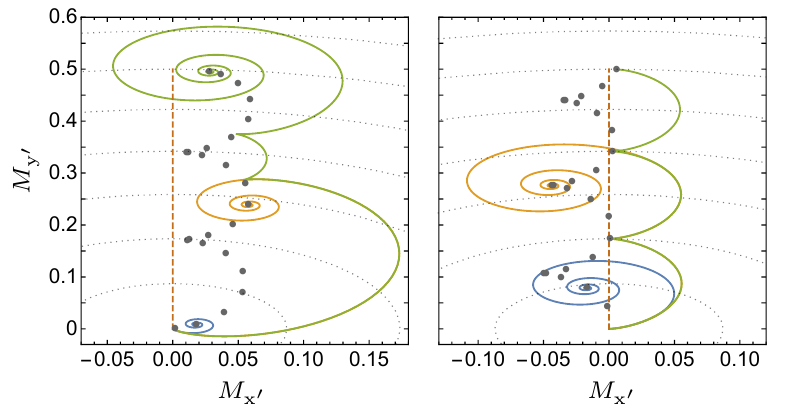} 
\caption{Simulations of small angle tips generated by a voltage-driven RL circuit ($\tau=T$) with 
$B_1= B_\omega/\nu_\pi$ (here $\nu_\pi =9$), $\delta B=0$, and $\phir =0^\circ$, for start angles $\phis =\phid=0^\circ$ (left) and $\phis =\phid=90^\circ$ (right). 
The filled circles indicate the components of the transverse magnetization in the rotating frame of the final terminus following 
rectangular pulses of rf voltage of increasing durations. The voltage pulse durations are integer multiples of $T/16$, up to a maximum of $3T/2.$ The full trajectories (solid lines) are shown for durations of $T/8$ (blue), $3T/4$ (orange), and $3T/2$ (green) only. 
Of these, only the case $t_\theta=3T/2$ with $\phis=90^\circ$  (green line, right graph) does not induce transients and thus 
has the trajectory expected for the corresponding rectangular $B_1$ pulse. 
The dashed lines are the RWA trajectories. The dotted lines indicate constant values of the polar angle in 5$^\circ$ steps.
}
\label{small_angle_LR}
\end{figure}

A further consideration regarding start phase and pulse duration comes when one contemplates driving an untuned coil with a voltage source as discussed in Section~\ref{sec:phasetrans}. In this case, it is best that one chooses both $t_\theta=n\,(T/2)$ and $\phis=90^\circ$ in order to achieve a true rectangular $B_1$ pulse and avoid transients at either end of the pulse, which otherwise have a large impact on small angle tips. Simulations comparing trajectories and terminus points for different pulse durations and start phases are shown in Fig.~\ref{small_angle_LR}. In general, due to the presence of transients, these results are quite different from the expectations for rectangular $B_1$ pulses both within and outside the RWA as shown in Figs.~\ref{fig:startPhase} and~\ref{small_angle_traj}. 

\section{Experimental system and methods: Key points}
\label{sec:experimental} 

This brief section contains key points and definitions needed for self-consistent reading of the remainder of the article. They are extracted from a fully developed Sec.~4 given in SM. 

\vspace{1mm}
\textit{Setup} -- All experiments were performed in a home-made low-field MRI system. They were performed at either a lower field of 0.8~mT ($f_{\mathrm{low}}$=25.7~kHz) using laser-polarized $^3$He gas enclosed in a sealed glass cell and polarized in-situ or at a higher field of 1.97~mT ($f_{\mathrm{high}}$=83.682~kHz) using thermally polarized water samples. 

Four different rf coils were used in the experiments (see Appendix~C in SM). 
We refer to the coils by the following acronyms: Sm and Sm$_{\mathrm{lowZ}}$ for two \uline{small} transmit coils (the latter being a low impedance version of the former, suitable for operation at $f_{\mathrm{high}}$); Lg for a \uline{large} transmit coil; and PU for the detection or \uline{pick up} coil. The Sm and Sm$_{\mathrm{lowZ}}$ coils were mainly used in this work and were left untuned. The Sm coil was only ever used at $f_{\mathrm{low}}$.
The Lg coil was used in some experiments and was tuned with a series capacitor when operating at $f_{\mathrm{high}}$.

\begin{figure}[htb]
\centering
\includegraphics[trim=0 0 0 0, clip=true, width=0.95\columnwidth]{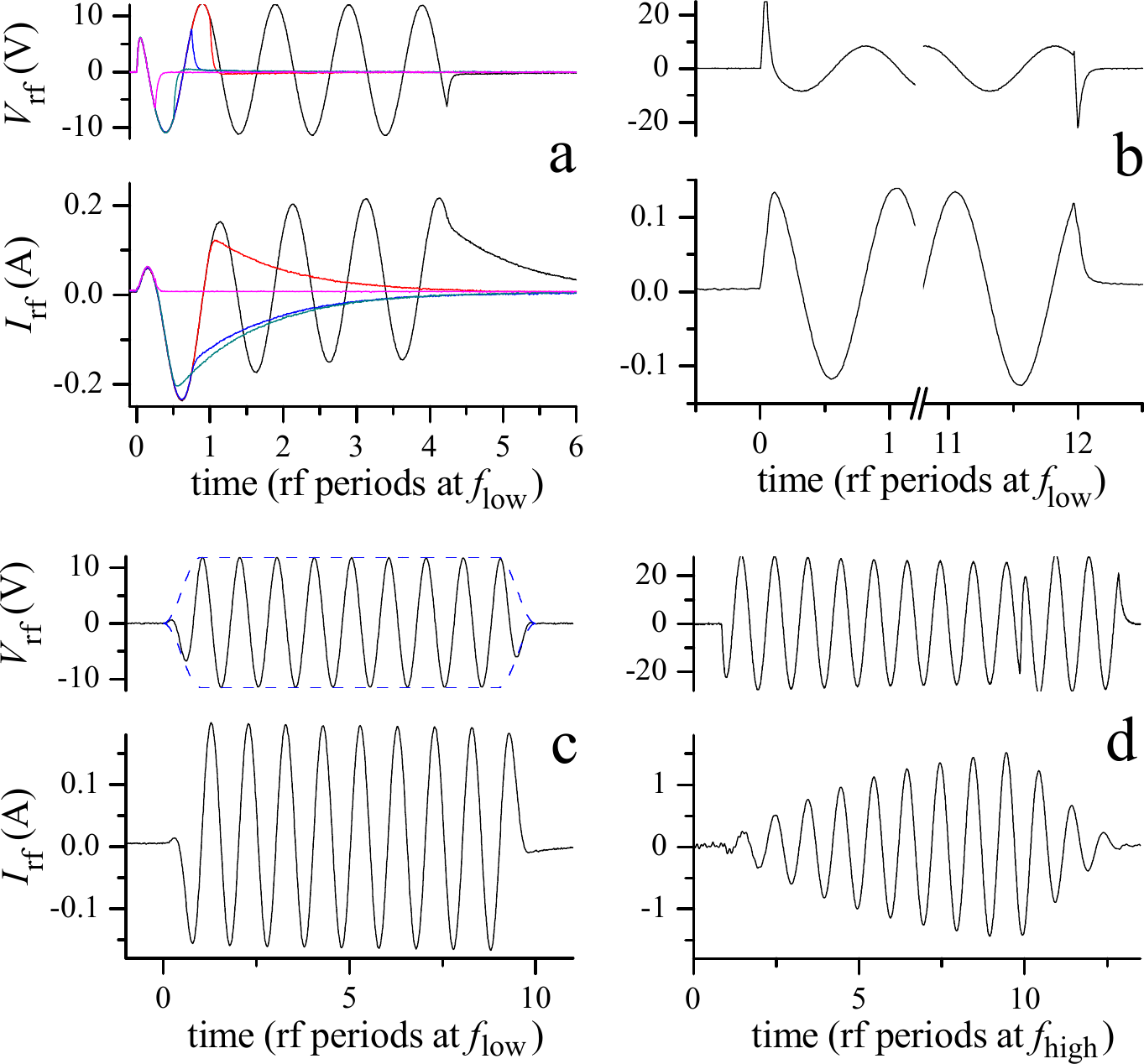} 
\caption{Driving voltages and resulting currents (upper and lower traces in each panel) recorded with the home amplifier and the Lg coil for different pulse shapes. a to c: untuned coil at $f_{\mathrm{low}}$. a: Rectangular pulses of various durations ($T$/4, $T$/2, 3$T$/4, $T$, and 17$T$/4) yield current responses with different transients (here, $\phis \approx -30{^\circ }$). b: Current transients were alleviated using suitable voltage bursts at the start and end of the pulse. c: A shaped pulse with 1-period-long cosine start and stop and a 8-period-long plateau nearly eliminated current transients. d: The coil was series-tuned at $f_{\mathrm{high}}$ and the current could progressively reach large values (note the scale difference). Current decay was accelerated by the voltage inversion \cite{Hoult1979}.}
\label{Fig-shapes}
\end{figure}

An Apollo Tecmag console was used to manage rf excitation pulses, static field shifts, spoiling gradients, and for data acquisition. 
The rf excitation current, which is proportional to $B_1$, was driven in the transmit coil using a home-built amplifier (100-W-peak, $Z=2$~$\Omega $). In some cases, explicitly noted in the text, a commercial pulsed NMR amplifier (250-W-peak, $Z=50$~$\Omega$) was employed. Experimentally-determined RL circuit time constants of the various coil and rf amplifier combinations are given in Table~C.1 in SM. 
Figure~\ref{Fig-shapes} displays four examples of driving voltages at the output of the amplifier and corresponding currents in the Lg coil in situations of interest in actual experiments. 
In Fig.~\ref{Fig-shapes}a simple rectangular 
pulses of rf voltage were applied to the untuned coil, as modeled in Sec.~\ref{sec:phasetrans} and Fig.~\ref{fig:currents}. In Fig.~\ref{Fig-shapes}b short periods with maximum driving voltages at the beginning and end of the pulse were used to reach the targeted current values at a high rate, approximating current ``jumps'' in the untuned coil. In Fig.~\ref{Fig-shapes}c a shaped voltage pulse of the form of Eq.~\ref{CPC_piecewise} 
was used to make smooth start and stop during the first and last rf periods of the pulse. Figure~\ref{Fig-shapes}d displays the response of the tuned coil to a voltage pulse shaped as suggested by Hoult~\cite{Hoult1979} to minimize transient times.

\vspace{1mm}
\textit{Data acquisition} -- The building blocks of all NMR measurements used here to determine tip angles are highlighted in Fig.~S.2 in SM 
for laser-polarized gas and for thermally-polarized water samples. Detuning was typically achieved through a shift $\delta B$ in the static field, but rf frequency shifts were sometimes applied. Working with fixed rf frequency was found to be imperative when using tuned coils as well as convenient for NMR sequence programming, as timings could be set based on a fixed rf period. 

When laser-polarized $^3$He gas was used, tip angles close to zero or $\pi$ hardly reduced the magnitude of the prepared longitudinal magnetization, and many tip angle measurements could be performed sequentially before the gas needed to be polarized again. 
When a water sample was used, the much lower available magnetization made operation at higher field more convenient and signal averaging mandatory. 

\vspace{1mm}
\textit{Data reduction} -- Recorded FID signals were processed for each recording to determine the following: initial magnitudes, from which tip angles $\theta$ were assessed; initial phases; and beat frequencies, from which differences $\Delta f$ between rf and Larmor frequencies were inferred. Details on the data reduction methods for experiments using $^3$He gas and water samples are given in Sec.~4.3 in SM. 

\section{Experimental results and discussion}
\label{sec:results}

All results reported here are for experiments performed starting from longitudinal magnetization. This was achieved via (near-)equilibrium magnetization in water samples (corresponding to magnetization at the north pole) or from hyperpolarized $^3$He gas samples (corresponding to magnetization at the north or south pole). 
Sections~\ref{results:RectangularUntuned} to~\ref{results:ShapedTuned} focus on the conditions required to generate accurate $\pi$-radian tips for a variety of pulse shapes using untuned and tuned coils. Results are reported in terms of $\beta$, which is the experimentally determined Bloch-Siegert shift $\delta B_\mathrm{p}$ scaled to an appropriate pulse-shape dependent shift (e.g.\ via Eq.~\ref{BS_shift_general} for rectangular pulses or Eq.~\ref{BSCW-max} for shaped pulses). Section~\ref{results:SmallTip} presents results for small tip angles achieved with rectangular and shaped voltage pulses using an untuned coil.

\subsection{Rectangular \texorpdfstring{$\bm{\pi}$} {Pi}-pulses using untuned coils}
\label{results:RectangularUntuned}

\begin{figure}[H]
\centering
\includegraphics[trim=0 0 0 0, clip=true, width=0.95\columnwidth]{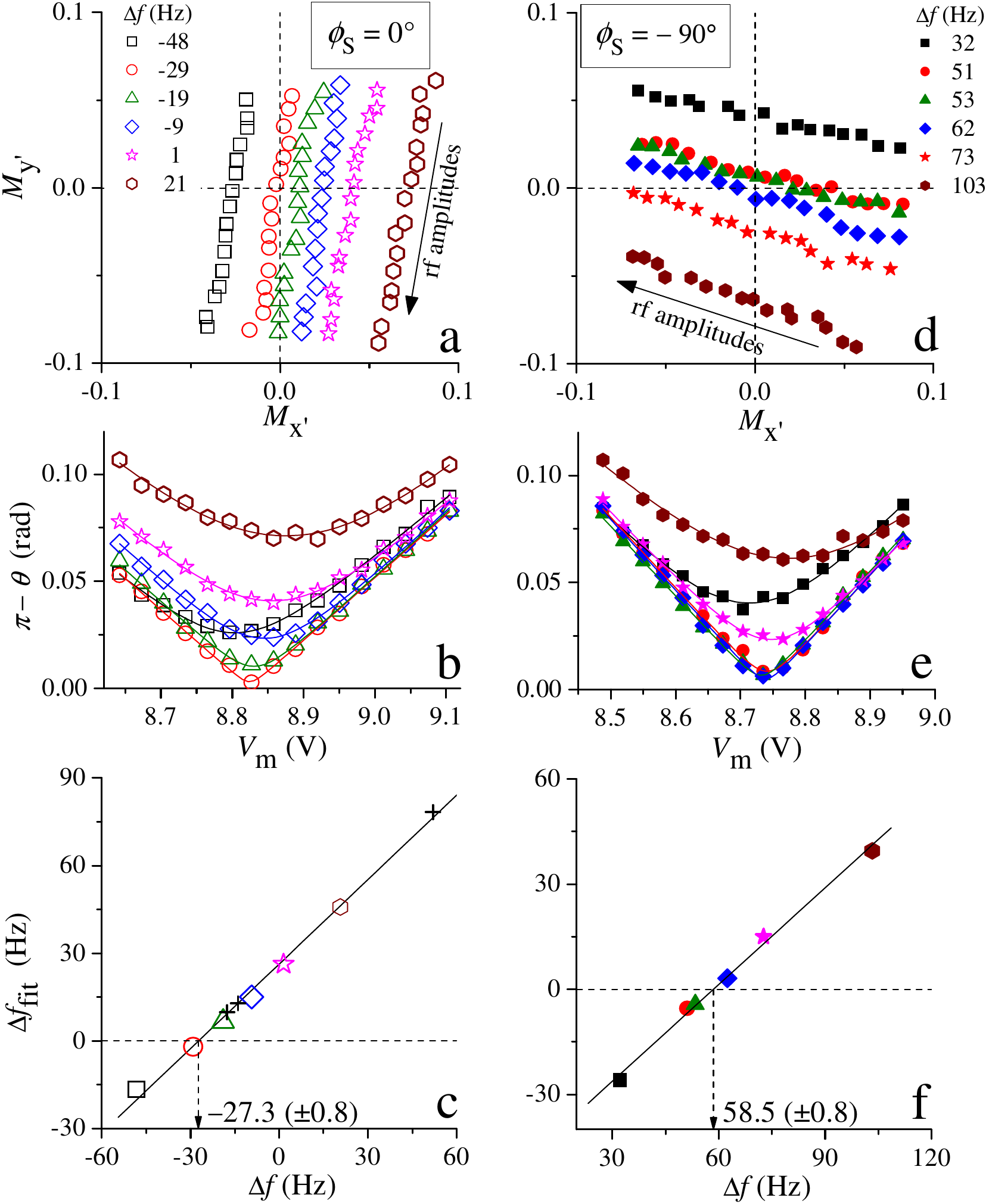} 
\caption{Reversal of longitudinal magnetization with 10-period-long rectangular pulses (tip angles $\theta \approx \pi$) for $\phis = 0^\circ$ and $-90^\circ$. a, d: Data points are transverse coordinates of tipped unit magnetization obtained for various rf pulse amplitudes $V_\mathrm{m}$. Each symbol type corresponds to a fixed value of the rf pulse detuning $\Delta f$ (see legends). b, e: The same data are used to plot the tip angle deviations from $\pi$ versus $V_\mathrm{m}$ (with corresponding symbol). The solid lines are fits used for each value of $\Delta f$ to evaluate the minimum deviation, characterized by an effective detuning $\Delta f_{\mathrm{fit}}$ (see text). c, f: Plots of the variations of $\Delta f_{\mathrm{fit}}$ with $\Delta f$, used to accurately evaluate the values of $\Delta f$ needed to achieve exact $\pi$-pulses, i.e. the Bloch-Siegert shifts associated with these pulses. The $+$ symbols in graph c correspond to additional data sets not displayed in graphs a and b for clarity. In all graphs, error bars are smaller than symbol sizes.}
\label{Fig-JSJ10p}
\end{figure}

Series of measurements were performed 
for 
$B_1$ pulses of approximate rectangular shape, 
with various durations $\nu _\pi T$, start phases, and detunings from resonance. 
Pulses lasting an integer number of rf periods were used, with durations ranging from $\nu _\pi =4$ to $\nu _\pi=30$ periods. Start phases were controlled via phase offsets of the rf carrier at fixed pulse timings.
They ranged from $\phis = -90^\circ$, giving sine-shaped currents, to $\phis = 0^\circ$, with near-jumps at the pulse start and end, giving cosine-shaped currents (see Fig.~\ref{Fig-shapes}b). 
For each set of pulse shape parameters, $\Delta f$ and $B_1$ (characterized by the driving voltage amplitude $V_\mathrm{m}$, see Eq.~\ref{eq:Vrf}) were varied by small steps in a 2D-acquisition sequence, 
sufficient to cover parameter space in the vicinity of the exact $\pi$-pulse
\footnote{This rather painstaking approach, purposefully taken here to provide a thorough investigation of the expected behavior outside the RWA, can in practice be replaced by a simpler $\pi$-pulse calibration method as discussed in Sec.~\ref{Sec6}.}  as per Fig.~\ref{fig:fine_tune}. Examples of data obtained for $\nu _\pi =10$ with $^3$He gas samples are displayed in Fig.~\ref{Fig-JSJ10p} for $\phis = 0^\circ$ (left panels a to c) and $\phis = -90^\circ$ (right panels d to f).

Each series of points for the two sets of measurements in Figs.~\ref{Fig-JSJ10p}a and \ref{Fig-JSJ10p}d represents the transverse components of the unit magnetization vector at its terminus 
in the vicinity of the targeted pole of the Bloch sphere as a function of $V_\mathrm{m}$ for a given $\Delta f$. 
As expected from Fig.~\ref{fig:fine_tune}, they lie on near-parallel straight lines in each figure, 
with directions rotated by the change in $\phis$ 
between the two sets of measurements. For each set, the minimum distance to the targeted pole is obtained for a non-zero detuning $\Delta f$, with strikingly different signs and values for the two start phases (see legends). 

Figures \ref{Fig-JSJ10p}b and \ref{Fig-JSJ10p}e display the tip angle errors $\pi - \theta$ for the same data points as functions of $V_\mathrm{m}$. A simple geometrical model for travel on the Bloch sphere in the RWA predicts that data sets are expected to lie on hyperbolas when $\pi - \theta \ll 1$ (Eq.~B.6 in SM). 
Each curve in Figs.~\ref{Fig-JSJ10p}b and \ref{Fig-JSJ10p}e is obtained from a non-linear least squares fit of the 3-parameter fitting function in Eq.~B.7 in SM 
to a data set, which yields: (i) an optimal value $V_\mathrm{fit}$ of the driving voltage amplitude for closest approach to the targeted pole, 
(ii) an asymptotic slope parameter $s_\mathrm{fit}$ for the hyperbolas asymptotes (found to be close to 1, which provides a consistency check of tip angle calibrations), and (iii) an effective detuning linked with the minimum deviation from the targeted pole, $\left|\Delta f_{\mathrm{fit}}\right| = (\pi - \theta )_{\mathrm{min}}f/4\nu_\pi $ ($f$ is the rf frequency).

Figures \ref{Fig-JSJ10p}c and \ref{Fig-JSJ10p}f compile the variations of the fitted detunings $\Delta f_{\mathrm{fit}}$ with the detunings $\Delta f$ from resonance (the signs of $\Delta f_{\mathrm{fit}}$ are deduced from the trajectories in Figs~\ref{Fig-JSJ10p}a and \ref{Fig-JSJ10p}d). The data in each panel are found to scale linearly with a slope close to 1, as expected. The detunings $\Delta f$ needed to exactly reach the targeted pole ($\Delta f_{\mathrm{fit}}=0$) are the Bloch-Siegert shifts inferred from the sets of measurements using this robust data reduction protocol. They are found to differ from the CW value of the Bloch-Siegert shift 
(16 Hz for these experimental conditions) and to strongly depend on the start phase in the pulse, as expected from the analytic and numerical predictions (Eq.~\ref{beta_rect} and Sec.\ref{sec:IdealRectPulses}). 
Appropriately scaling the measured Bloch-Siegert shifts yields the prefactor $\beta$ of Eq.~\ref{BS_shift_general}.

\begin{figure}[htb]
\centering
\includegraphics[trim=0 0 0 0, clip=true, width=0.99\columnwidth]{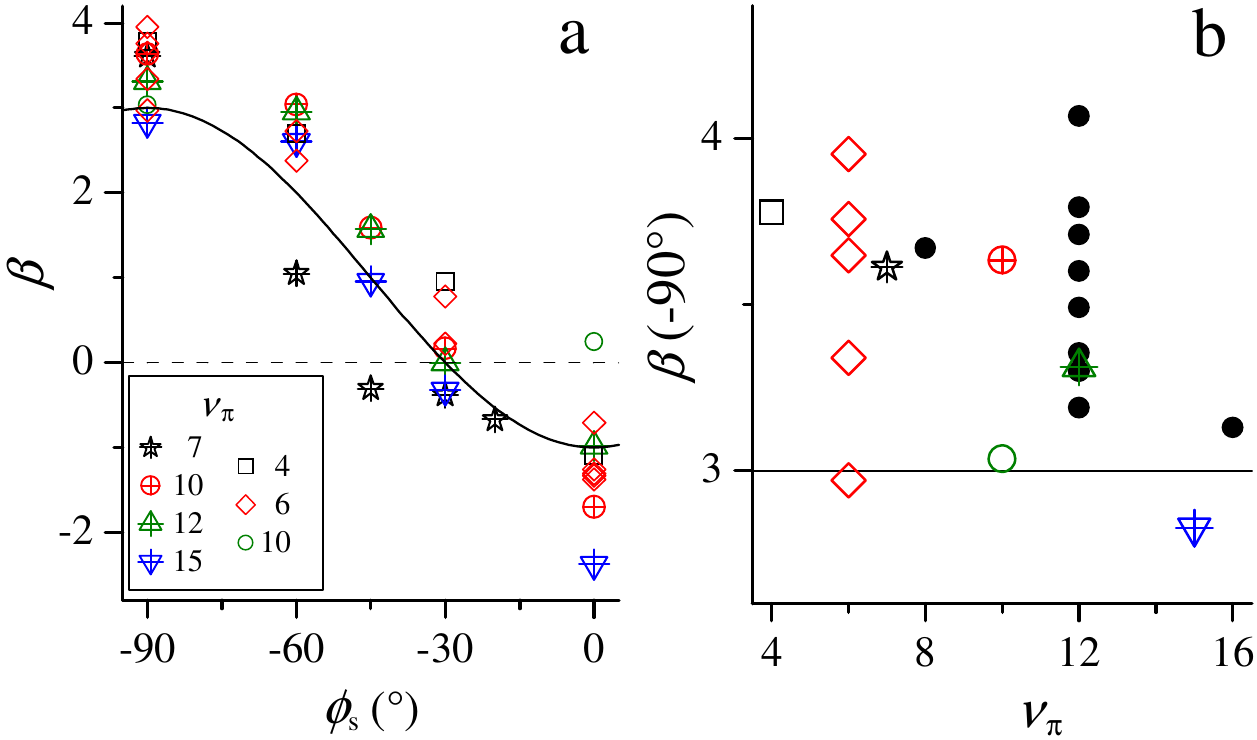} 
\caption{
$\beta(\phis)$ for rectangular current pulses. 
a: Theoretical curve (solid line) and experimental results (symbols) inferred as in Fig.~\ref{Fig-JSJ10p} from experiments with $^3$He at $f_{\mathrm{low}}$. Experiments were performed for various pulse lengths (see the legend) using the Sm and Sm$_{\mathrm{lowZ}}$ coils (open and crossed symbols, respectively).
 b: Results for $\phis =-90^{\circ}$ plotted versus pulse length (note the expanded vertical scale). $^3$He data (same data and symbols as in Fig.~\ref{Fig-JSJall}a) 
and water data (solid symbols) were obtained at $f_{\mathrm{low}}$ and $f_{\mathrm{high}}$, respectively.}
\label{Fig-JSJall}
\end{figure}

Values of $\beta$ determined for rectangular $B_1$ 
$\pi$-pulses of various durations and start phases are compiled in Fig.~\ref{Fig-JSJall}. 
Figure~\ref{Fig-JSJall}a displays results obtained at $f_{\mathrm{low}}$ with $^3$He as a function of the start phase $\phis$. Figure~\ref{Fig-JSJall}b displays all results obtained for $\phis =-90^\circ$ 
with $^3$He (at $f_{\mathrm{low}}$) and water samples (at $f_{\mathrm{high}}$). 
The expected variation of $\beta$ with $\phis$ (Eq.~\ref{beta_rect} and solid line in Fig.~\ref{Fig-JSJall}a) is globally obeyed by the experimental results, as well as the independence of $\beta$ on the pulse length at fixed $\phis$ (Fig.~\ref{Fig-JSJall}b), consistent with the $\nu _\pi ^2$ scaling of the Bloch-Siegert shifts as per Eq.~\ref{BS_shift_general}. 
The significant scatter of the results and their deviations from the expected values are attributed to deviations of the actual currents in the rf coil from perfect rectangular pulses. When numerical simulations are performed using experimentally recorded currents as inputs, the results of the simulations and of the corresponding experiments agree within experimental uncertainties (which are smaller than the symbol sizes in Figs.~\ref{Fig-JSJall}a and \ref{Fig-JSJall}b). 
Simulations show that the finite rise and fall times of the current jumps and the additional initial and final deviations from ideal rectangular pulses, which result from the slew-rate and bandwidths limitations of the rf amplifiers (e.g.\ Fig.~\ref{Fig-shapes}b), fully account for differences between experimental results and theoretical expectations in Fig.~\ref{Fig-JSJall}. 
Indeed, minor changes in voltage parameter settings between repeat experiments -- which were made to obtain satisfactory current pulse shapes via visual minimization of initial and final artifacts -- resulted in the observed variation of $\beta$ at nominally fixed pulse 
 shape ($\nu_\pi$ and $\phis$).

\begin{figure}[htb]
\centering
\includegraphics[trim=0 0 0 0, clip=true, width=0.85\columnwidth]{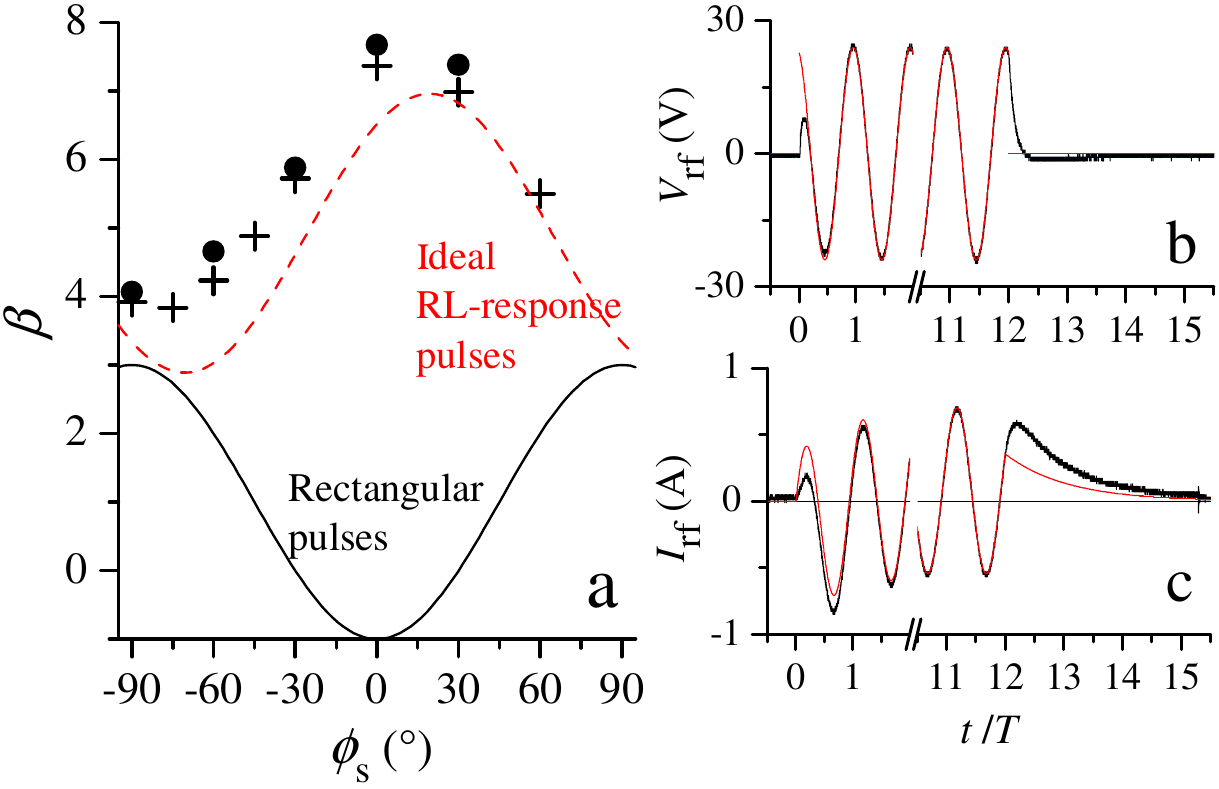} 
\caption{
$\beta(\phis)$  for rectangular voltage pulses with $\nu _\pi =12$.
a: Experimental results ($\bullet $) differ strongly from expectations for rectangular current pulses (solid line). 
Values inferred from numerical simulations for currents expected from the ideal RL-circuit response to a rectangular voltage pulse (dashed line) and for the actual currents recorded in the experiments ($+$) display a similar enhancement near $\phis=0$. 
b and c: An example of voltage and current pulses for $\phis = -30^{\circ}$. The noisy traces (black) are recorded data, the overlaid smooth traces (red online) are model functions with instant voltage jumps (b) and the corresponding computed current response (c) used in the simulations which yield the dashed line (a).}
\label{Fig-LRwater}
\end{figure}

Given the sensitivity of $\pi$-pulse accuracy to the rise and fall of current shapes, we have not tried to make Bloch-Siegert shift measurements involving current jumps at the higher frequency since pulse shape distortions would only become more pronounced. On the contrary, we performed measurements for rectangular pulse shapes of the applied voltage, such as one may be tempted to use in an elementary approach of NMR. Indeed, the resulting current pulse shapes are strongly affected by the transient response of the coil-amplifier system (Sec.~\ref{sec:phasetrans} and Fig.~\ref{Fig-shapes}a). Results obtained with water samples at $f_{\mathrm{high}}$ with the Sm$_\mathrm{lowZ}$ coil
are displayed as solid symbols in Fig.~\ref{Fig-LRwater}a as a function of start phase $\phis $, which is defined, as before, to be $-90^\circ$ for a sine-like current in the coil.
The most striking feature is the strongly enhanced Bloch-Siegert shifts (up to 8-fold compared to $\beta =1$, i.e. to the CW shift) 
which are observed for this excitation scheme. The slow initial and final pulse distortions account for most of the difference with rectangular current pulses (dashed line versus solid line). Additional initial and final deviations from ideal rectangular voltage pulses (Fig.~\ref{Fig-LRwater}b), which result (as above) from imperfections of the rf amplifiers, further enhance the computed shifts and almost fully account for the observations ($+$ symbols).

\subsection{Shaped \texorpdfstring{$\bm{\pi}$} {Pi}-pulses using untuned coils}
\label{results:ShapedUntuned}

\begin{figure}[htb]
\centering
\includegraphics[trim=0 0 0 0, clip=true, width=0.75\columnwidth]{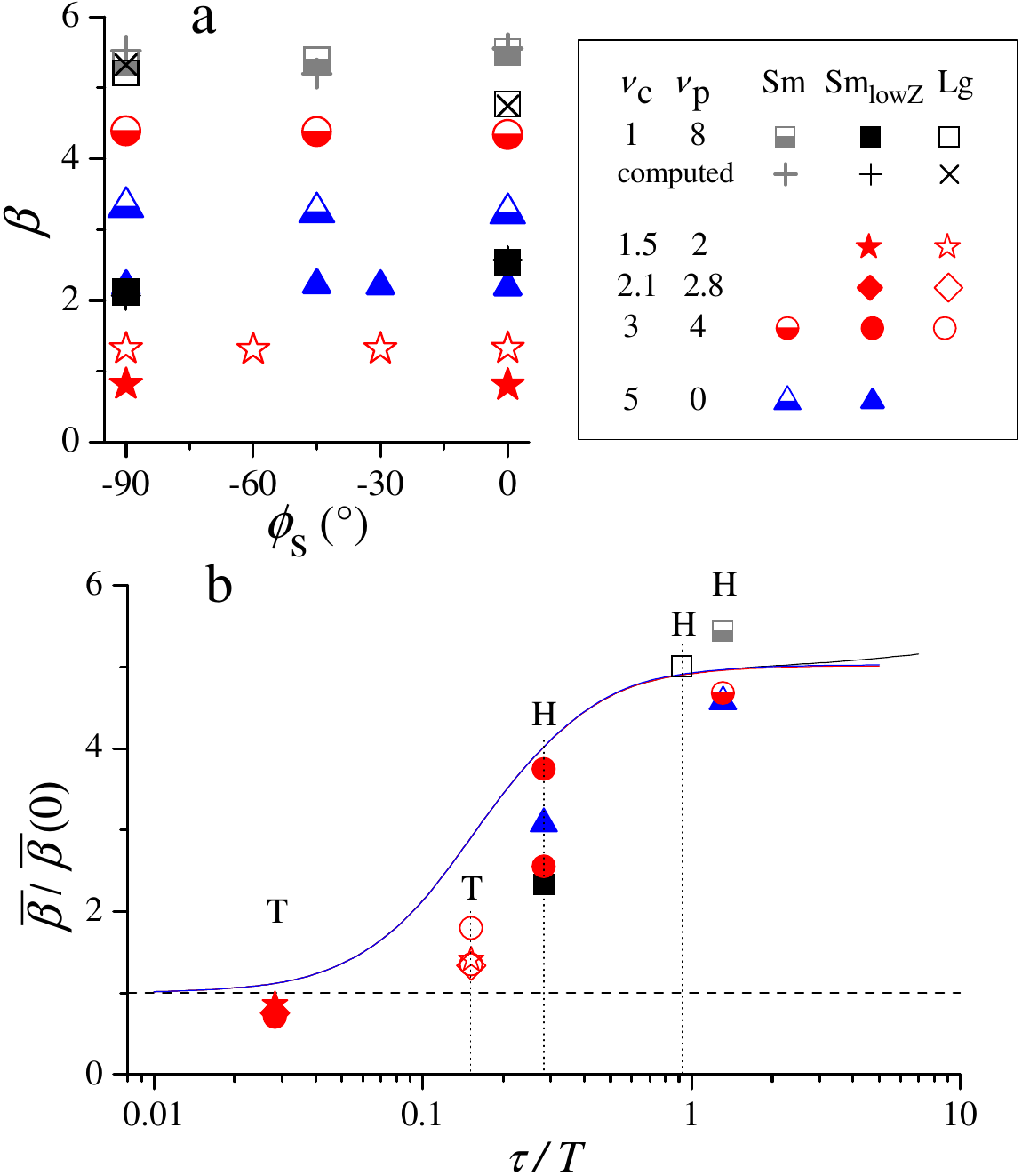} 
\caption{
$\beta$ for a variety of voltage pulse shape parameters and rf coils as indicated in the legend. 
 The symbol colors (online) correspond to the families of pulse shapes as discussed in the text. All experiments were performed with $^3$He at $f_{\mathrm{low}}$. 
a: Results plotted versus start phase for select series only (for the sake of clarity).
Computed results from recorded currents for the pulse with $\nu_{\mathrm{c}}=1$ and $\nu_{\mathrm{p}}=8$ are shown as well.
 b: Averaged values of $\beta(\phis)$ for all series plotted versus reduced time constant of the coil/amplifier combination (vertical dotted lines as visual guides) and further distinguished by whether the Tomco (T) or home-built (H) amplifier was used. 
 The results are scaled to the average expected value of $\beta(\tau \!=\!0$) (as per Fig.~\ref{Fig-CPC-Volts}). The solid lines are spline interpolations of averages generated for each of the three families of simulated data in Fig.~\ref{Fig-CPC-LR}. 
 }
\label{Fig-CPCall}
\end{figure}

Experiments were performed with $^3$He at $f_{\mathrm{low}}$ for three different families of voltage pulse shapes corresponding to those presented in Sec.~\ref{sec:shaped} :
(i) $\nu_{\mathrm{c}}=1$ and $\nu_{\mathrm{p}}=8$, (ii) different constructions of the ratio $\nu_{\mathrm{c}} \! : \! \nu_{\mathrm{p}} = 3 \! : \! 4$, and (iii) $\nu_{\mathrm{c}}=5$ and $\nu_{\mathrm{p}}=0$. Different combinations of rf coil and amplifier were employed here in order to explore a broad range of $\tau$ values. The prefactor $\beta$ was extracted as per Sec.~\ref{sec:shaped}. The results for individual experiments are compiled in Fig.~\ref{Fig-CPCall}a and show that as expected from simulations $\beta$ is largely 
independent of $\phis$ with all other parameters being fixed.  As a check, values of $\beta(\phis)$ were computed from recorded currents for a few of the experiments and show good agreement, here again, with the measured results.

The data were further processed by taking the average value $\bar \beta$ of $\beta(\phis)$ for each series of experiments in Fig.~\ref{Fig-CPCall}a (i.e.\ fixed values of $\nu_{\mathrm{c}}$ and $\nu_{\mathrm{p}}$ and a given coil) and normalizing it to the corresponding average value of $\beta$ for ideal shaped current pulses (i.e.\ $\tau=0$), which are the solid symbols in Fig.~\ref{Fig-CPC-Volts}. The resulting data are plotted in Fig.~\ref{Fig-CPCall}b versus the reduced time constant for the coil/amplifier combination used for that series. Also shown here are theoretical curves of $\bar \beta(\tau) / \bar\beta(0)$
determined from spline interpolation of the simulated data in Fig.~\ref{Fig-CPC-LR}. The results for $\nu_{\mathrm{c}}=3$ and~5 are indistinguishable at this scale over the entire range. At large $\tau/T$, the result for $\nu_{\mathrm{c}}=1$ breaks from the other two, owing again to the small effect of $\phis$ at $\tau=0$ for this case (see inset in Fig.~\ref{Fig-CPC-LR}). Overall, there is global support for the expected universal scaling here, with experiments confirming that for untuned coils driven by shaped voltage pulses (i) the relevant CW values of the Bloch-Siegert shift ($\beta \approx 1$) is required for $\tau/T \ll 1$
and (ii) a strong enhancement of the shift (up to five times) is required for $\tau/T > 0.5$.

\subsection{Shaped \texorpdfstring{$\bm{\pi}$}{Pi}-pulses using a tuned coil}
\label{results:ShapedTuned}

Series of experiments performed 
with doped water 
at $f_{\mathrm{high}}$ were made using the
tuned Lg transmit coil for faster, more energy-efficient tipping. In such
cases, rectangular current pulses cannot be achieved simply (see, however,
Ref.~\cite{Nacher2020}) and delays and transients in the rf field amplitude $B_1$ 
occur on time scales of the order of $QT$, where $Q$ is the quality factor of the coil 
(e.g.\ $Q_{\mathrm{Lg}} \approx15$, see~Appendix C in SM). 
The impact of these effects was explored here for shaped voltage pulses with (i) cosine-like rise and fall times as defined in Sec.~\ref{sec:shaped} and (ii) fast rise and fall times as shown in Fig.~\ref{Fig-shapes}d.

\begin{figure}[ptb]
\centering
\includegraphics[trim=0 0 0 0, clip=true, width=0.87\columnwidth]{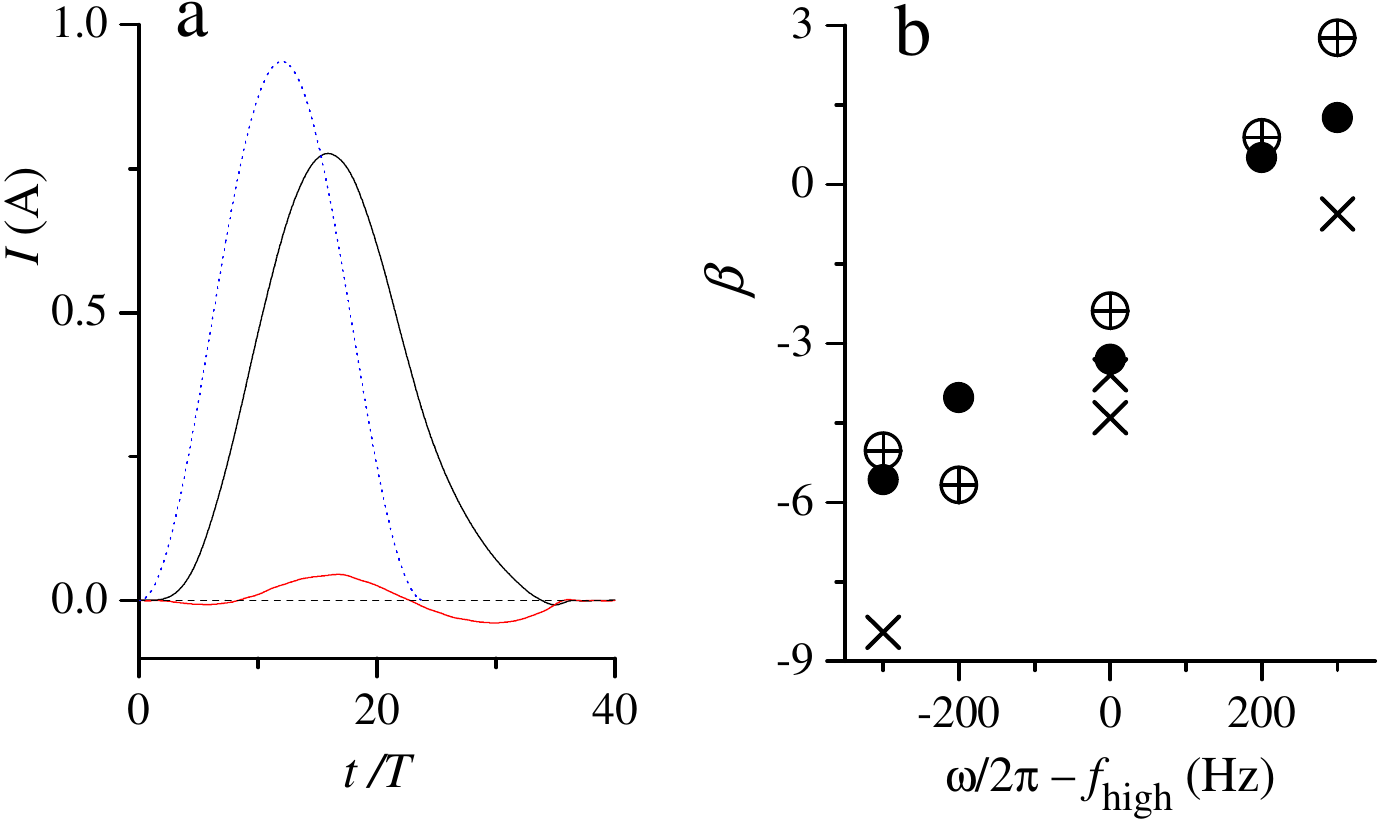} 
\caption{
Recorded coil current (a) and $\beta$ values (b) for shaped pulses ($\nu_{\text{c}}=12$ and
$\nu_{\text{p}}=0$)  
at or close to $f_{\text{high}}$. 
a: In-phase (black solid line) and out-of-phase (red solid
line) components of the demodulated rf current during a transmit pulse. The
blue dotted line is the scaled amplitude of the driving voltage, with the same
integral as the current magnitude.
b: Values of $\beta$ versus offset of the rf frequency inferred from NMR data (solid circles), computed from associated  coil currents following each NMR data set (crossed circles),  and computed from coil currents  recorded in quick succession of one another (crosses). 
}
\label{Fig:CPCtuned}
\end{figure}

For the first case, we chose \textit{bell-shaped} voltage pulses with $\nu_{\mathrm{c}}=12$ and
$\nu_{\mathrm{p}}=0$ (i.e., no plateau between the rise and fall).
These  yield $\phi_{\mathrm{s}}$-independent negative Bloch-Siegert shifts
of the order of $-130~\mathrm{Hz}$.
 If the expression of Eq.~\ref{BSCW-max} is used to scale the data to a relevant CW value of the Bloch-Siegert
shift, prefactors $\beta\approx-3.5$ are inferred. They strongly differ from the positive values of
$\beta$ (ranging from 1 to 5) obtained with untuned transmit coils
(Fig.~\ref{Fig-CPCall}). 
Using recorded rf currents to compute Bloch-Siegert shifts does indeed yield negative $\beta$ values consistent with
those found in the experiments. Spectral analyses of the
currents reveal that their average frequencies differ from $f_{\mathrm{high}}$. 
They also exhibit phase transients as shown in the  example of
Fig.~\ref{Fig:CPCtuned}a
in which the two plotted quadrature components are obtained by numerical demodulation of a recorded current with
suitable low-pass filters (62-points FFT smoothing for 192 ns sampling).
 As expected, the current magnitude (not shown) is
significantly delayed and distorted with respect to the driving voltage, and phase changes are evidenced. Such delays and phase transients
(or glitches) in the response of a tuned RLC circuit can normally be computed,
and their effects on NMR dynamics are well documented
\cite{Hoult1979,Ellett1971,Barbara1991,Greer2021}, but the non-linear response
of the noise-blocking diodes, for instance, makes the computation of NMR dynamics
from recorded currents plausibly more reliable for our experiments.

Figure \ref{Fig:CPCtuned}b displays $\beta$ values inferred from sets of
experiments, each performed at fixed rf frequency $\omega/2\pi$,
fixed rf amplitude, and variable static field shift $\delta B$. 
To explore the sensitivity  of $\beta$ to potential
differences  between rf frequency and coil resonance (as might occur due to mistuning or thermal drifts, say), 
$\omega/2\pi$ was offset by various amounts from $f_{\mathrm{high}}$, 
while the coil remained tuned to this value.\footnote{
In practice, it was  $\sim 50$~Hz below $f_{\mathrm{high}}$, which was the best  tuning that could be achieved with reasonable effort.  We kept $f_{\mathrm{high}} = 83.682$~kHz as our nominal operating frequency, however, as it provided a convenient rf period for the 100~ns time resolution of our NMR console \cite{Nacher2020}.
}
The $\beta$ values were found to depend strongly 
on the offset of the rf frequency in a consistent way between NMR-inferred data (solid
symbols) and values computed from recorded currents. More regular behaviours were
obtained for currents recorded within seconds of each other (crosses in Fig.~\ref{Fig:CPCtuned}b), as opposed to delays of tens of minutes (the typical acquisition time for each NMR data set),
which is attributed to possible temperature-induced drifts of the coil
inductance and resonance frequency.

\begin{figure}[htb]
\centering
\includegraphics[trim=0 0 0 0, clip=true, width=0.97\columnwidth]{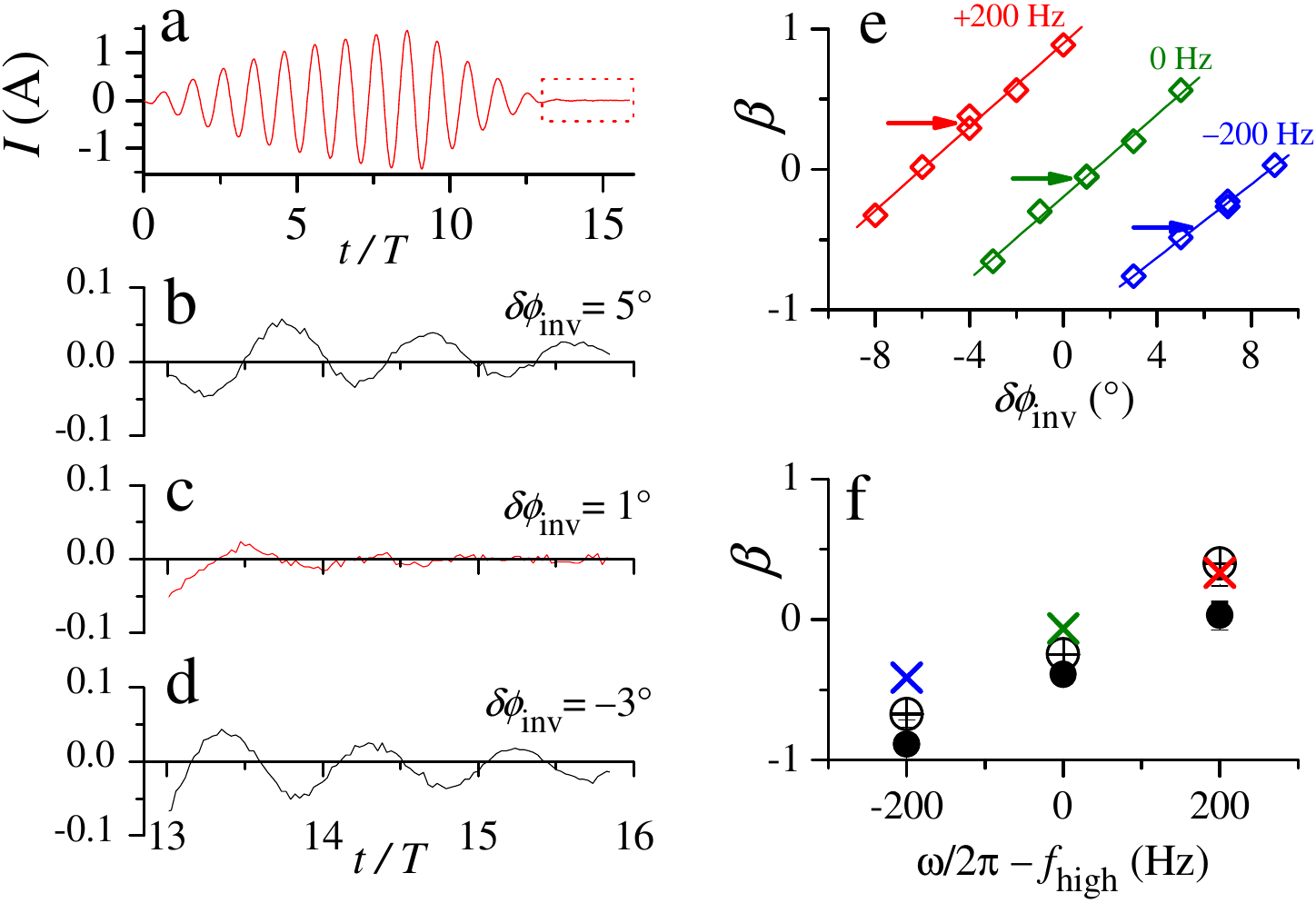} 
\caption{
Recorded coil current (a-d) and $\beta$ values (e,f) for fast shaped voltage pulses 
at or close to $f_{\mathrm{high}}$. a: Current for
well-adjusted parameters at $f_{\mathrm{high}}.$ b to d: Expanded views of
unwanted current ring-downs for different values of the additional phase shift
$\delta\phi_{\mathrm{inv}}$ (see text).  
e: Values of $\beta$ versus $\delta\phi_{\mathrm{inv}}$ computed from coil currents  recorded in quick succession of one another.  Data are color-coded (online) with respect to rf frequency offset as indicated on the graph. The value of $\beta$ associated with the optimal  $\delta\phi_{\mathrm{inv}}$ at each frequency is indicated by an arrow. 
f: Values of $\beta$ versus offset of the rf frequency inferred from NMR data (solid circles), computed from associated  coil currents following each NMR data set (crossed circles),  and determined from the optimal $\delta\phi_{\mathrm{inv}}$ values in e (color-coded crosses). 
}
\label{Fig:RKtuned}
\end{figure}

For the second type of shaped pulses  explored here, fast current rise and fall times were obtained by applying a constant voltage amplitude for a duration of  $9T$ followed by an inverted (i.e.\ $\pi$-shifted)  voltage  at 75\% amplitude for $4T$
in order to \textit{dump} the initially stored energy~\cite{Hoult1979}.  
Such straightforward adjustment can normally be achieved only at
resonance~\cite{Hoult1979}, but a small additional phase shift $\delta
\phi_{\mathrm{inv}}$ of the inverted voltage during the \textit{dumping} step can otherwise be used to fully
cancel the rf current. An example of rf current driven at  $f_{\mathrm{high}}$
(close to resonance) is shown in Fig.\ \ref{Fig:RKtuned}a, with  expanded views of the
currents after the nominal end of the pulse (the region in the red dotted box)
stacked in Figs.\ \ref{Fig:RKtuned}b to d for three values of $\delta
\phi_{\mathrm{inv}}$.  Optimal values of $\delta\phi_{\mathrm{inv}}$ were visually
assessed for each rf frequency: $4^\circ$,
 $1^\circ$, and $-4^\circ$ for frequency shifts of $-200$, 0, and $+200\;\mathrm{Hz}$ from
$f_{\mathrm{high}},$ respectively.  

Figures \ref{Fig:RKtuned}e and \ref{Fig:RKtuned}f display $\beta$ values inferred from sets of experiments performed around $f_{\mathrm{high}}.$ 
In keeping with our approach thus far, we compute $\beta$ here using an effective duration associated with the voltage pulse shape.  In this case, given that the inverted section of the pulse would ostensibly reverse the direction of nutation, one must subtract the duration of the second part of the pulse from that of the first, weighting both with respect to rf amplitude.  The result is an effective pulse duration of $6T$.   This definition is not unique, of course, and one could even define an effective pulse duration from the resulting current shape, if so desired. But while any particular definition will affect the magnitude of the resulting $\beta$ value, it will not alter its sign nor  its sensitivity to an rf frequency offset or a non-optimal choice of $\delta\phi_{\mathrm{inv}}$.

Figure~\ref{Fig:RKtuned}e shows a consistent  linear dependence of $\beta$ on $\delta\phi_{\mathrm{inv}}$ for different rf frequency offsets, as computed from recorded coil currents.  
Each arrow corresponds to the optimal value of $\delta\phi_{\mathrm{inv}}$ determined from a second-order polynomial fit of  the area of the amplitude of the unwanted current ring down.   
The associated  $\beta$ values are plotted in Fig.~\ref{Fig:RKtuned}f (crosses) along with  data determined from NMR experiments.   
As in Figure \ref{Fig:CPCtuned}b, $\beta$ is found to  depend strongly on the offset of the rf frequency for NMR-inferred data (solid symbols) as well as for values computed from recorded currents (all other symbols). 
Overall, these results highlight the sensitivity of $\beta$ to frequency offsets, as well as to fine details of the pulse that might otherwise be ignored, such as small amounts of current ring-down.

\subsection{Small tip angles using an untuned coil}
\label{results:SmallTip}

\begin{figure}[htb]
\centering
\includegraphics[trim=0 0 0 0, clip=true, width=0.95\columnwidth]{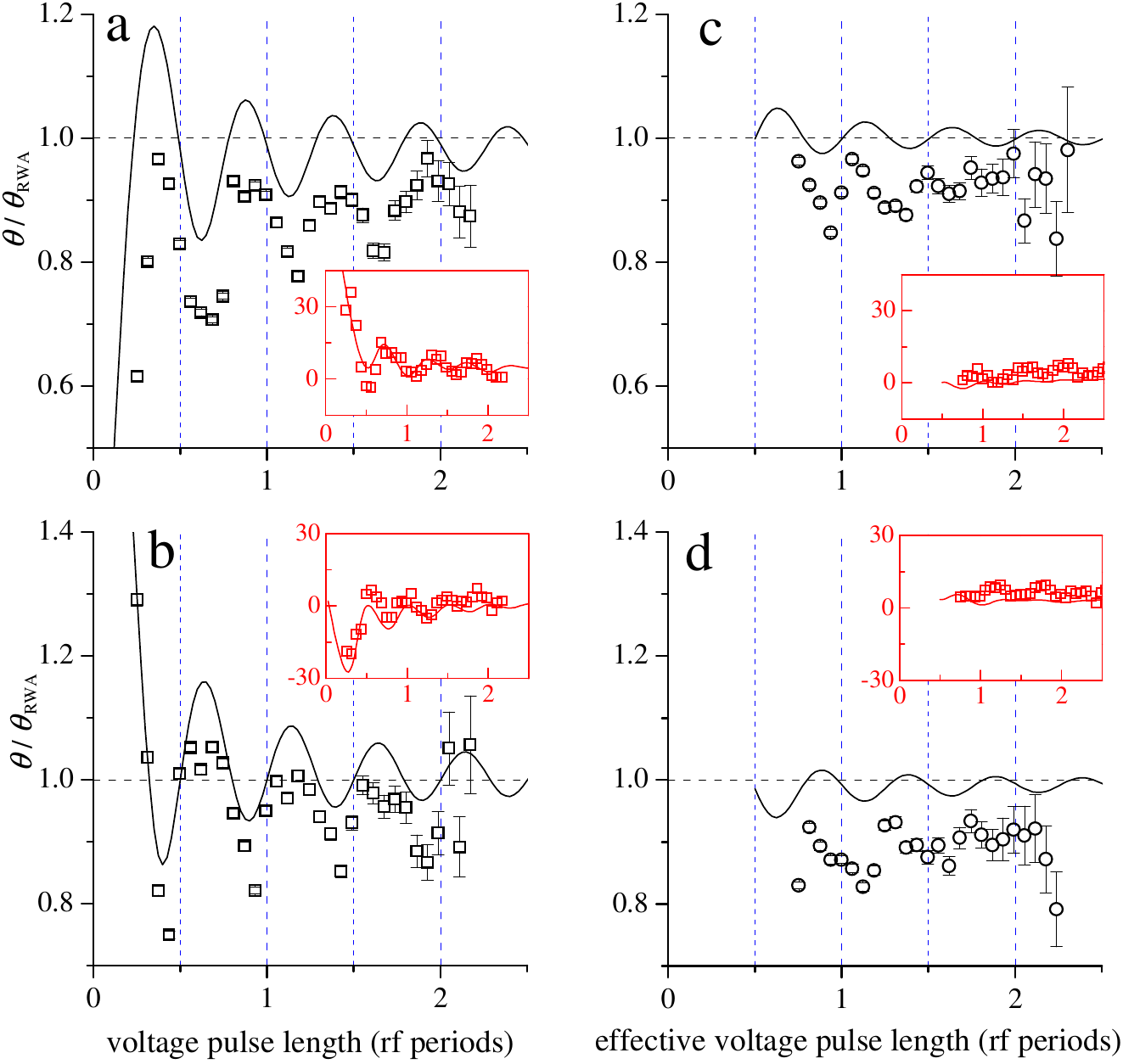} 
\caption{
Relative tip angle $\theta / \theta_{\mathrm{RWA}}$ versus pulse length for rectangular voltage pulses (a, b) and shaped voltage pulses (c, d) in an untuned coil for start phases $\phis = 0^\circ$ (a, c) and $-90^\circ$ (b, d). 
Insets: Phase of the transverse magnetization (in degrees) immediately following the pulse.
NMR data are indicated by the symbols and the numerical results by solid lines.
All experiments were performed with $^3$He at $f_{\mathrm{low}}$ with $\delta B=0$ and rf amplitude set by $\nu_\pi=9$ (a,b) or \ $\nu_{\mathrm{c}}=0.5$ and $\nu_{\mathrm{p}}=8.5$ (c,d).
 The minimum effective pulse length achieved experimentally for the shaped pulses (i.e.\ $\nu_{\mathrm{c}}=0.5$ and $\nu_{\mathrm{p}}=0.25$) is 0.75 rf periods. The error bars increase with pulse length because data were taken in quick succession from small to larger tip angles, reducing the available $^3$He magnetization along the way as described in Sect.~4.2 in SM. 
As a result, data with pulse lengths $\ge 2.25$ rf periods yielded little signal and have been clipped.
 }
\label{Fig-smallflips}
\end{figure}

A final set of experiments using the untuned Sm coil was performed with $^3$He at $f_{\mathrm{low}}$ to explore the generation of small tip angles starting from longitudinal magnetization. Both rectangular and shaped voltage pulses were employed, with $B_1 = B_\omega / 9$ (i.e. $\nu_\pi=9$). Pulse durations ranged from $0.25$ to $2.25$ rf periods for the rectangular pulses, which in the limit of the RWA, and in the absence of phase transients, would ideally produce tip angles $\theta_{\mathrm{RWA}} = 5^\circ - 45^\circ$. 
For the shaped pulses, a constant value of $\nu_{\mathrm{c}}=0.5$ was used while $\nu_{\mathrm{p}}$ was varied from $0.25$ to $2.25$, giving a range of effective pulse durations from 0.75 to 2.75 rf periods, corresponding to 
$\theta_{\mathrm{RWA}} = 15^\circ - 55^\circ$. 
Data for relative tip angle $\theta / \theta_{\mathrm{RWA}}$, as well as the phase of the transverse magnetization following the tip, are presented in Fig.~\ref{Fig-smallflips} for the cases $\phis = 0^\circ$ and $-90^\circ$. Theoretical curves that incorporate voltage pulse shape and the RL response of the circuit (as discussed in Sections~\ref{sec:shaped} and~\ref{sec:smalltips}) are included. 

The results show that, as expected, deviations from RWA values vary with a half-period cycle. This provides experimental confirmation that in the low field/frequency regime one cannot arbitrarily shorten an rf pulse that has been calibrated for a larger tip angle and expect to achieve the desired proportional reduction in tip angle, nor maintain the expected phase of the transverse magnetization. 
 The imperfect agreement between 
 numerical and experimental results
 for $\theta / \theta_{\mathrm{RWA}}$ is, we believe, due to the limitation in the accuracy of our tip angle calibration described in Sect.~4.2 in SM. 
This issue notwithstanding, the results also demonstrate that shaped voltage pulses lead to smaller deviations from RWA values (Fig.~\ref{Fig-smallflips}c and~d).
For either pulse type, numerical results show that $\theta / \theta_{\mathrm{RWA}} =1$ for 
$t_\theta=nT/2$, which again suggests that keeping pulse durations to integer multiples of the rf half-period is a convenient strategy for achieving desired tip angles when using untuned coils.

Data for repeat measurements with other values of the start phase -- including $\phis=30^\circ$ for which $\beta =0$ for rectangular pulses -- are not presented here but exhibit similar behavior to that displayed in Fig.~\ref{Fig-smallflips}. This confirms that the oscillations are due to the presence of the CR component of the rf field and, as discussed in Section~\ref{sec:smalltips},
 are further modified by transients in the case of the rectangular pulses. The effect of detuning (i.e. $\delta B \neq 0$) was also explored, and as expected has little to no impact on the magnitude of small angle tips but does modify the phase of the resulting transverse magnetization. 

\section{Summary and conclusions}
\label{Sec6}

We have presented an extensive -- but certainly not exhaustive -- exploration of the many combined effects  that one must consider in order to generate accurate tip angles with pulsed linear rf fields outside the RWA.  These include detuning (via the static field or rf frequency) and pulse characteristics such as  phase, amplitude, shape,  and  transients.   Numerical simulations (via time integration of the Bloch equation) and low-field NMR experiments provided access to a broad range of behaviors within this large parameter space.

It was found numerically, and confirmed by first-order analytic treatments 
(Appendix~A in SM), 
that the trajectory and terminus of the magnetization on the Bloch sphere achieved by 
rectangular $B_1$ pulses is greatly influenced by the start and end phase of the pulse. 
For example, when starting from the north pole a terminus at 
exactly the south pole can be achieved for a pulse duration $\nu_\pi T=nT/2$ only if $\phis=30^\circ$.  For all other values of the start phase, one will miss the south pole.
This runs counter to conventional RWA experience, 
which would suggest that any such pulse of appropriate area (i.e.\ amplitude times duration) can be gated at anytime to give the desired tip angle.

The desired terminus can still be achieved for any start phase and pulse duration, if one applies the appropriate shift  from resonance (Bloch-Siegert shift) and an accompanying change in rf amplitude. 
In this manner, 
one compensates for inaccuracies 
that were induced by operating outside the RWA by employing these two controls to nullify the transverse magnetization 
at the terminus.  
For rectangular $\pi$-pulses starting with transverse magnetization,  simulations also showed that  the same optimal parameters (field shift and rf amplitude) needed to achieve an exact tip from north to south pole will 
eliminate any possible over/undershoot of the transverse plane. 

In this work, we chose to cast  the static field shift  in terms of the original Bloch-Siegert shift for CW excitation multiplied by a prefactor $\beta$. For weak rectangular $B_1$ pulses lasting an integer multiple of the rf half-period (i.e.\ $\nu_\pi =n/2 \gg 1$), simulation and calculations show that $\beta(\phis) =1-2\cos2\phis$, resulting in shifts that lie anywhere between $-1$ and $3$ times that of the CW value of the shift. In general, $\beta$ depends on many parameters including pulse duration, pulse amplitude, pulse shape, rf circuit impedance, and whether the rf coil is untuned or tuned.  For example, simulations for shaped current pulses (comprising a cosine-like rise and fall that smoothly transitions with a central plateau region) showed that $\beta$ depends on $\phis$ and oscillates about unity for short rise/fall times, whereas for long rise/fall times $\beta$ depends  weakly  on the aspect ratio of the pulse and becomes $\phis$-independent. For similarly shaped pulses of rf voltage  driving an untuned coil, the same trend in $\beta (\phis)$ is seen as a function of rise/fall time, but there is a significant enhancement (and eventual saturation) in the value of $\beta$ as the coil time constant increases. 

Experimental results were  presented here primarily in terms of $\beta$.  Overall, there is good agreement with expectations.  Deviations and scatter are attributable to the  challenge of generating sufficiently accurate replications of current and voltage waveforms in practice (owing to amplifier slew-rate and bandwidth limitations) and reveals the  unexpected, and somewhat surprisingly large, sensitivity of the results to the fine details of coil currents and rf fields. Further refinement of experimental procedures may be needed for future applications of this work but was not crucial for this initial, exploratory study.

For rectangular pulses of rf current in an untuned coil, the form of  $\beta(\phis)$ given by Eq.~\ref{beta_rect} was confirmed, with values ranging from $-1$ to $+3$.  For rectangular pulses of voltage and untuned coils, the RL response of our particular coil/amplifier configuration significantly altered the start phase dependence of $\beta$, with a large positive enhancement near $\phis=0^\circ$.  For  shaped  pulses of voltage and an untuned coil, $\beta$ was found to be largely independent of $\phis$ but depended strongly on the coil circuit time constant.  For shaped voltage pulses and a  tuned coil, $\beta$ was again independent of $\phis$, but was highly sensitive to any offset between the rf drive frequency and the coil resonance, as well as the level of suppression of current ring-downs.

Finally, both simulations and experiments were performed using shortened pulses in order to explore the generation of small tip angles. Overall, the results  demonstrate that outside the RWA one cannot arbitrarily shorten an rf pulse that has been calibrated for a larger tip angle and expect to achieve the desired proportional reduction in tip angle.  Shaped pulses induced less deviation from such expectations than rectangular pulses. A robust strategy that avoids tip angle errors all together with the latter is to use pulse durations that are an integer multiple of the rf half-period along with a zero-current start, i.e.\ $\phis=\pm 90^\circ$.

Several practical considerations arise from the work presented here.  First, given that the complex trajectories on the Bloch sphere that occur outside the RWA are not readily deduced from usual NMR experience, simulations should be performed to gain insight.  One should also record coil currents during an experiment (or monitor $B_1$ directly with a search coil, say) and use these in simulations to check results and further guide experimental design.   
Next, we found it advantageous to fix the rf frequency and detune  via a shift in the static field. This approach provides a fixed time base in terms of the rf period $T$ that is convenient for  managing pulse sequence timings.  Lastly, in regard to $\pi$-pulse calibration (of any pulse shape),  it is much less time-consuming to begin with a 1D parameter search to find the rf amplitude that minimizes the FID signal induced by a pair of identical pulses (yielding a $2\pi$-rotation) for which the terminus back near the north pole is largely independent of the applied field shift $\delta B$.  
The only  requirement is that no significant free evolution of the magnetization close to the south pole occurs in the rotating frame during the time gap between the pulses, a condition that is automatically obtained at resonance ($\delta B=0$). Once the correct $\pi$-pulse amplitude has been found, finding the field shift $\delta B_\mathrm{p}$ needed to exactly invert magnetization becomes a second simple 1D minimization task. In a similar way, $\pi/2$-pulse parameters can be determined by first using a series of four pulses to find the optimal $B_1$ amplitude, then using a series of two pulses to determine $\delta B_\mathrm{p}$.

Future work will include the numerical and experimental study of composite pulses, and series of such pulses, applied using standard or phase-gradient rf coils for various low-field applications. A detailed study of the impact of the concomitant longitudinal component of the rf field on tip angle is in progress. It is believed to give rise to artifacts observed in low-field TRASE images~\cite{Nacher2018,Bidinosti2018a}  and may be of similar concern here as is the presence of static concomitant gradient fields in conventional low-field  imaging~\cite{Norris1990,Ullah2009,Nieminen2010}. 


\section*{Acknowledgments}
We gratefully acknowledge the support of the University of Winnipeg, the Natural Sciences and Engineering Research Council of Canada, L'\'{E}cole normale sup\'{e}rieure, and the Centre National de la Recherche Scientifique of France.
\balance
\putbib
\clearpage
\end{bibunit}
\nobalance 

\twocolumn [         
\begin{@twocolumnfalse}   
\begin {center}
{\LARGE Generating accurate tip angles for NMR outside the rotating-wave\\
\vspace{0.2cm} 
 approximation: Supplemental material}\\
\vspace{0.6cm} 
{\large Christopher P. Bidinosti, Genevi\`eve Tastevin, Pierre-Jean Nacher}
\end{center}
\hrule
\vskip 0.25cm
This supplemental material to the article ``Generating accurate tip angles for NMR outside the rotating-wave approximation'' comprises one section (Sec.~4) and the three appendices (A to C) of the originally submitted version of the article that were removed from the body of the article for the sake of brevity. The body of the article was slightly edited to accommodate this removal, with a significantly reduced form of the full Sec.4 given here.

\vskip 0.25cm
\hrule
\vskip 1cm
\end{@twocolumnfalse}  
]   

\renewcommand\thefigure{S.\arabic{figure}}    
\renewcommand{\theequation}{S.\arabic{equation}}
\setcounter{figure}{0}    
\setcounter{equation}{0}
\setcounter{section}{3}
\begin{bibunit}
\section{Experimental system and methods: Detailed description}
\label{sec:experimentalSM}  
\subsection{Setup}
\label{sec:SetupSM}

All experiments were performed in a home-made low-field MRI system as described in Refs.~\cite{Nacher2020,Safiullin2013}.
Independent additional windings on the coil formers 
were used to produce a uniform static-field shift $\delta B$.
Three sets of gradient imaging coils were used for first-order field shimming, yielding a cell-volume-dependent half-life of FID signals, $T_2^{\ast }$, of order 100~ms at 2.5~mT. The MRI system resides inside a 0.5-mm thick copper Faraday cage~\cite{Bidinosti03}, which reduced interference noise over the range of operating frequencies (25--110~kHz).

 Experiments were typically performed at either a lower field of 0.8~mT ($f_{\mathrm{low}}$=25.7~kHz) using laser-polarized $^3$He gas or at a higher field of 1.97~mT ($f_{\mathrm{high}}$=83.682~kHz) using a thermally polarized water sample. The $^3$He gas (400~mbar, in a sealed glass cell made via a careful in-house preparation protocol)
 was polarized in-situ. An ac high voltage at roughly 2~MHz was applied to the two interleaved electrodes wound on the outer surface of the cell (see Fig.~\ref{fig-exp2}a, also Ref.~\cite{Baudin11}) to generate the weak discharge needed for metastability-exchange optical pumping (MEOP)~\cite{Gentile}. Forced air flow around the cell prevented significant heating due to the discharge. Pumping with 1--2~W of incident laser light typically yielded a steady-state nuclear polarization $P_{\infty}$=2\% with a build-up time constant $T_\mathrm{b}$=100~s. 
 For convenience, the MEOP process was left on during NMR experiments. This was similar to operating in an effective magnetic field of 7500~T with 
 a longitudinal relaxation time $T_\mathrm{b}$. The water sample filled a 8.3-cm-long section of a 20.4~mm i.d.\ cylindrical glass tube and was doped with 
 CuSO$_4$ (2.8~mM concentration),
 giving $T_1$=$T_2 \approx $200~ms.

Four different rf coils were used in the experiments. Details on their construction and characteristics are provided in Appendix~\ref{sec:RFcoil}. We refer to the coils by the following acronyms: Sm and Sm$_{\mathrm{lowZ}}$ for two \uline{small} transmit coils (the latter being a low impedance version of the former, suitable for operation at $f_{\mathrm{high}}$); Lg for a \uline{large} transmit coil; and PU for the detection or \uline{pick up} coil. The Sm and Sm$_{\mathrm{lowZ}}$ coils were mainly used in this work and were left untuned. The Sm coil was only ever used at $f_{\mathrm{low}}$.
The Lg coil was used in some experiments and was tuned with a series capacitor when operating at $f_{\mathrm{high}}$.
 The PU coil was tuned with a parallel capacitor. 
Electronic feedback was used to actively broaden the detection bandwidth and accelerate ring-down in the detection circuit of emfs due to the rf pulse or $Q$-switching (see details in Appendix~\ref{sec:RFcoil} and in Sec.~3.4 of \cite{Kuzmin2020}).

An Apollo Tecmag console (low-frequency LF1 model), allowing for 100~ns time resolution of sequence events, was used to manage rf excitation pulses, preamplifier blanking, and relay control (through optocouplers, HCPL-2231, located inside the Faraday cage). 
 It was also used for data acquisition through a high-impedance, low-noise preamplifier (SR560, Stanford Research Systems). 
 The console gradient control unit was used to manage static field shifts, as well as spoiling gradients when needed by the measurement sequences.

\begin{figure}[H]
\centering
\includegraphics[trim=0 0 0 0, clip=true, width=0.98\columnwidth]{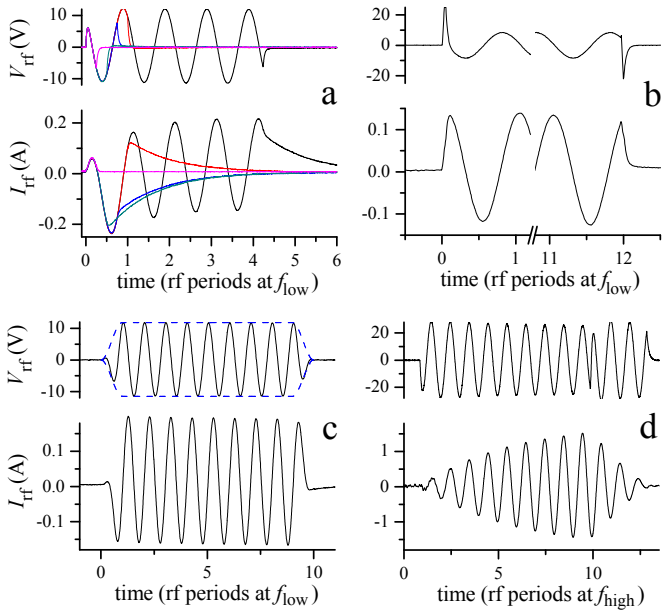} 
\caption{Driving voltages and resulting currents (upper and lower traces in each panel) recorded with the Lg coil for different pulse shapes. a to c: untuned coil at $f_{\mathrm{low}}$. a: Rectangular pulses of various durations ($T$/4, $T$/2, 3$T$/4, $T$, and 17$T$/4) yield current responses with different transients (here, $\phis \approx -30{^\circ }$). b: Current transients were alleviated using suitable voltage bursts at the start and end of the pulse. c: A shaped pulse with 1-period-long cosine start and stop and a 8-period-long plateau nearly eliminated current transients. d: The coil was series-tuned at $f_{\mathrm{high}}$ and the current could progressively reach large values (note the scale difference). Current decay was accelerated by the voltage inversion \cite{Hoult1979}.}
\label{Fig-shapesSM}
\end{figure}

The rf excitation current, to which $B_1$, is proportional, was driven in the transmit coil using a home-built amplifier (100-W-peak, $Z=2$~$\Omega $). In some cases, explicitly noted in the text, a commercial pulsed NMR amplifier (Tomco Technologies, BT00250-Alpha A, 250-W-peak, $Z=50$~$\Omega$) was employed. The current was monitored using a low-inductance 1-$\Omega $ resistor on the return line from the coil to the amplifier ground. Experimentally-determined RL circuit time constants of the various coil and rf amplifier combinations are given in Table~\ref{tab:Coils}.
Figure~\ref{Fig-shapesSM} displays four examples of driving voltages at the output of the latter amplifier and corresponding currents in the Lg coil in situations of interest in actual experiments. 
In Fig.~\ref{Fig-shapesSM}a simple rectangular 
pulses of rf voltage were applied to the untuned coil, as modeled in Sec.~2.3 and Fig.~3. In Fig.~\ref{Fig-shapesSM}b short periods with maximum driving voltages at the beginning and end of the pulse were used to reach the targeted current values at a high rate, approximating current ``jumps'' in the untuned coil. In Fig.~\ref{Fig-shapesSM}c a shaped voltage pulse of the form of Eq.~13 
was used to make smooth start and stop during the first and last rf periods of the pulse. Figure~\ref{Fig-shapesSM}d displays the response of the tuned coil to a voltage pulse shaped as suggested by Hoult~\cite{Hoult1979} to minimize transient times.

\subsection{Data acquisition}
\label{sec:DataAquisitionSM}
\begin{figure}[H]
\centering
\includegraphics[trim=0 0 0 0, clip=true, width=0.85\columnwidth]{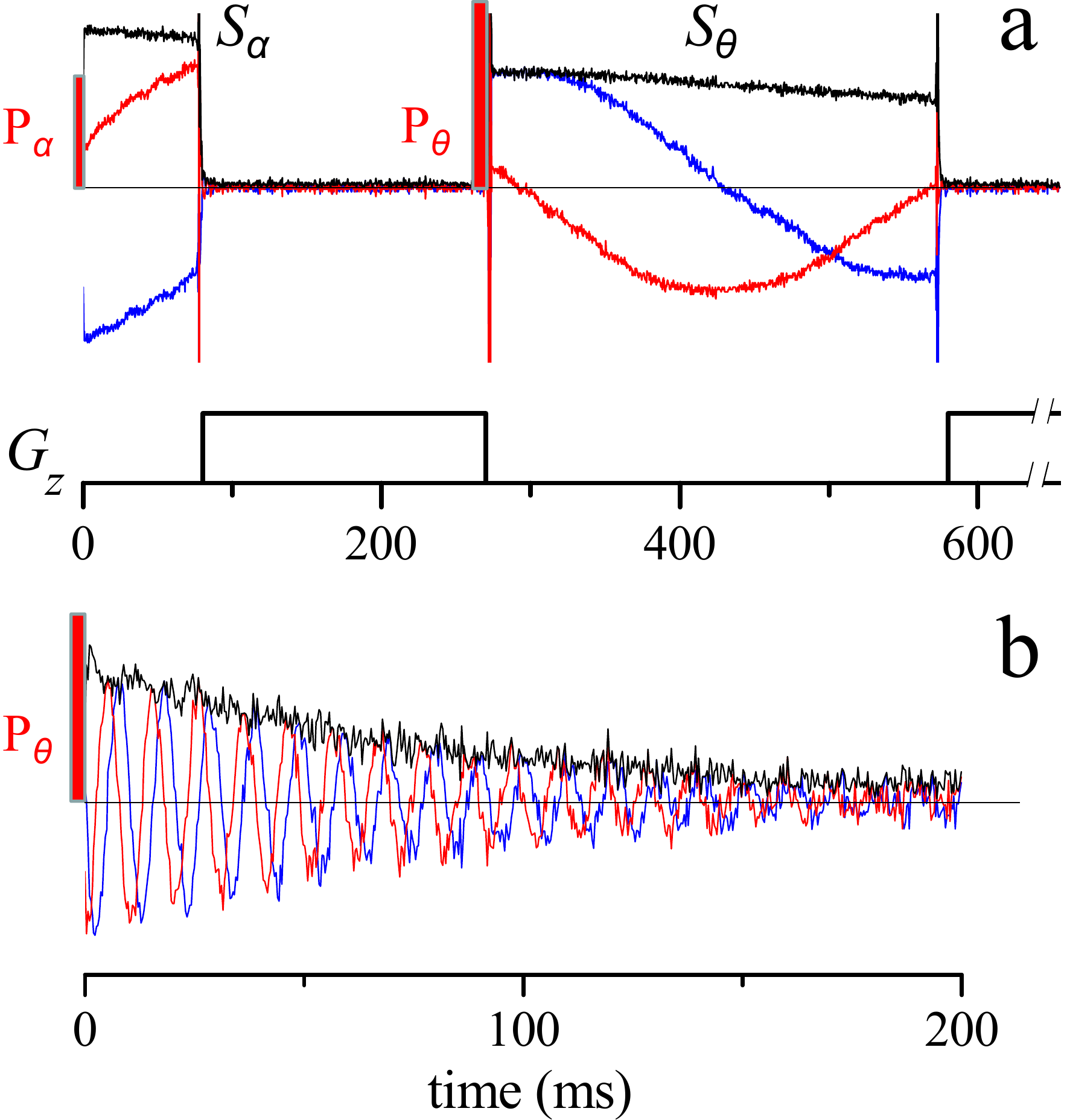} 
\caption{Elementary measurements and recorded signals are shown for two kinds of experiments. The blue and red oscillating traces (color online) are demodulated quadrature signals, the black traces are the corresponding signal magnitudes. a: Two pulses were used for laser-polarized gas (P$_\alpha$ is a fixed measurement pulse yielding a signal $S_\alpha$ proportional to the available magnetization; P$_\theta$ was the variable pulse under test yielding a variable signal $S_\theta$). Gradient pulses (in the z-gradient coil) were used to spoil transverse magnetization after each signal measurement ($G_z$=1.7~mT/m, $k=6.7\times 10^4$~m$^{-1}$, $D=0.8\times10^{-4}$ m$^2$/s, $T_\mathrm{d}=2.8$ ms). b: A single variable pulse was used for experiments with water samples. Sufficient time between repeated pulses was chosen (here, $T_\mathrm{R}$= 470~ms) so that a large fraction of the thermal polarization was recovered between pulses (Eq.~\ref{DR3}).}
\label{fig-seqHeH}
\end{figure}

The building blocks of all NMR measurements used here to determine flip angles are highlighted in Fig.~\ref{fig-seqHeH} with examples of recorded signals from laser-polarized gas and thermally-polarized water samples. 
Detuning was typically achieved through a shift $\delta B$ in the static field, but shifts of the rf frequency $f$ were sometimes applied. Working with fixed rf frequency (the former method) is imperative when using tuned coils. We also found this approach to be convenient for NMR sequence programming, as timings could be set based on a fixed rf period. The difference $\Delta f$ between the transmitted rf field frequency and the observed free precession frequency 
provides an accurate measurement of the detuning. For simplicity it is always reported here in units of Hertz, regardless of how the experiment was performed.

When laser-polarized $^3$He gas was used, flip angles close to zero or $\pi$ hardly reduced the magnitude of the prepared longitudinal magnetization, and many flip angle measurements could be performed sequentially before the gas needed to be polarized again (on a time scale of $T_\mathrm{b}$). Because of the variable available magnetization, a measurement pulse P$_\alpha$ was always used before the pulse under test P$_\theta$ was applied (Fig.~\ref{fig-seqHeH}a). Free induction decay signals were recorded following each pulse. With the long-lived signals achieved at 0.8~mT, gradient pulses were used to spoil the transverse magnetization, which rapidly decayed due to atomic diffusion in the gas. Measurements were performed every 800~ms for different values of the rf amplitude or duration of the P$_\theta$ pulse using 2D sequences without signal averaging ($N_\mathrm{R}$=1). The one-rf-period-long measurement pulse P$_\alpha$ was periodically calibrated by two methods: (1) monitoring signal decay when P$_\theta$ was inhibited ($T_1$ decay was checked to be negligible), and (2) using a $\pi /2$ P$_\theta$ pulse. Its amplitude was set to yield $\alpha=6.6^{\circ}$ from both calibration methods.

When a water sample was used, the much lower available magnetization made operation at higher field more convenient and signal averaging mandatory. Figure \ref{fig-seqHeH}b displays an example of recorded signal for $\theta=16^{\circ}$ and $N_\mathrm{R}$=128. The detection system was periodically calibrated by performing standard inversion-recovery measurements (P$_{\pi}$ - $T_\mathrm{IR}$ - P$_{\pi/2}$ - CPMG) with variable $T_\mathrm{IR}$ delays, which yielded the values of the maximum achievable signal magnitude $S_\infty $ and of the relaxation times $T_1$ and $T_2$ of the doped water sample.

\subsection{Data reduction}
\label{sec:DataReductionSM}
Recorded FID signals were processed for each recording to determine the following: initial magnitudes, from which tip angles $\theta$ were assessed; initial phases; and beat frequencies, from which $\Delta f$ values were inferred. Beat frequency fluctuations ($\pm $ a few Hz) were due to slow fluctuations of the background magnetic field. Initial signal magnitudes $S$ were conveniently inferred from time-dependent signal magnitudes $S(t)$ (black traces in Fig.~\ref{fig-seqHeH}) using parabolic fits over a suitable range, which phenomenologically combine $T_2$ and $T_2^{\ast}$ decays to lowest order. Similarly, numerically unwrapped phases were computed from the quadrature signals. Parabolic fits, phenomenologically accounting for beat frequencies and their lowest-order drifts, were used to infer the initial phases and beat frequencies. 

This simple data reduction method breaks down at low SNR, where Rician noise in magnitudes introduces significant bias and statistical uncertainties on phases and frequencies become large. This occurs whenever the tip angle $\theta $ is very close to 0 or $\pi $. The simple approach was therefore replaced in this case with a 2-step method.
First, the beat frequency was inferred from a high-SNR FID signal (e.g. $S_\alpha$ in 2-pulse experiments, Fig.~\ref{fig-seqHeH}a, or $S_\theta$ from a different value of $\theta $ in 1-pulse experiments, Fig.~\ref{fig-seqHeH}b). The quadrature signals, $P(t)$ and $Q(t)$, were combined to eliminate this beat, as if demodulation were performed at the Larmor frequency. Second, the resulting quadrature signals, $P'(t)$ and $Q'(t)$ were processed using parabolic fits, allowing the initial signal magnitude and phase, $S=[P^{\prime}(0)^2 + Q^{\prime}(0)^2 ]^{1/2}$ and $\varphi =\Arg[P^{\prime}(0) + i Q^{\prime}(0)],$
 as well as beat frequency correction, to be inferred.

The tip angle $\theta $ induced by P$_\theta$ from an initially longitudinal magnetization was computed using
\begin{equation}
\sin \theta = S_\theta /S_\mathrm{max},	
\label{DR1}
\end{equation}
where the maximum signal magnitude $S_\mathrm{max}$ which would be obtained for $\theta =\pi/2$ was assessed differently for $^3$He gas (for which longitudinal relaxation is overlooked) and water samples:
\begin{eqnarray}
S_\mathrm{max} = S_\alpha \cot \alpha 	& 	\mbox{for $^3$He (Fig.~\ref{fig-seqHeH}a)} \label{DR2} \\
S_\mathrm{max} = \frac{1-\cos\theta\; e^{-T_\mathrm{R}/T_1}}{1-e^{-T_\mathrm{R}/T_1}} S_\infty 	&	\mbox{for water (Fig.~\ref{fig-seqHeH}b).} \label{DR3}
\end{eqnarray}
In practice the correction factor which reduces $S_\mathrm{max}$ below $S_\infty $ in steady-state (Eq.~\ref{DR3}) is often close to 1 (for small tip angles) or weakly depends on $\theta $ (for $\theta \approx \pi$). Therefore the solutions of Eq.~\ref{DR1} for water were easily obtained iteratively from the data sets.

\appendix
\section* {APPENDIX}
\renewcommand\thefigure{A.\arabic{figure}}    
\renewcommand{\theequation}{A.\arabic{equation}}
\setcounter{figure}{0}    
\setcounter{equation}{0}
\section{Analytic solutions of  \texorpdfstring{$\bm{\beta(\phis)}$}{beta(phis)}}
\label{sec:appendixA}

Here we evaluate to lowest significant order the effect of Larmor detuning
and of the counter-rotating (CR) component of the rf field applied
during a rectangular pulse by taking suitable approximations in the doubly rotating
reference frame corresponding to the unperturbed nutation of the magnetization (i.e., no CR component, and no Larmor detuning).  As in Sec.~2.2, we consider the scenario where the rf angular frequency $\omega$ is fixed and subsequently determine the shift in static field $\delta B_{\rm p}$ away from the value $B_\omega = \omega / \left|\gamma\right|$ that is needed to achieve an accurate $\pi$-pulse.

Since the relevant details of the spin trajectory depend only on the start phase (Sec.~2.1),  it is natural here to let $\phi_r=0$ and  set $\phis = \omega \td$
via a time delay $0<\td<T/2$, where $T$ is the rf period and $\ts=\td$ is now the start time at which a rectangular pulse of the linear rf field is turned on  instantaneously.  
In the laboratory and rotating frames (via Eqs.~1 and~2, respectively) this rf field takes the form
\begin{eqnarray}
\bm B_1(t) \!\!\! 	& = & \!\!\! 	B_1 \cos \omega t \ \bm{\hat{x}} \\ 
\label{B1_lab_appendix}
\bm{\mathcal{B}}_1^+(t) \!\!\! 	&=&	\!\!\! 	 \tfrac{1}{2} B_1 \left[ (1+\cos 2\omega t) \  \bm{\hat{x}'}  
+\sin 2\omega t  \ \bm{\hat{y}'}  \right] \, .
\label{B1_rot_appendix} 
\end{eqnarray}
We do not consider variations in the rf amplitude here, so $  B_1 = B_\omega / \nu_\pi$ is the nominal RWA value used throughout.

\begin{figure}[tbh]
\centering
\includegraphics[trim=0 0 0 0, clip=true, width=\columnwidth]{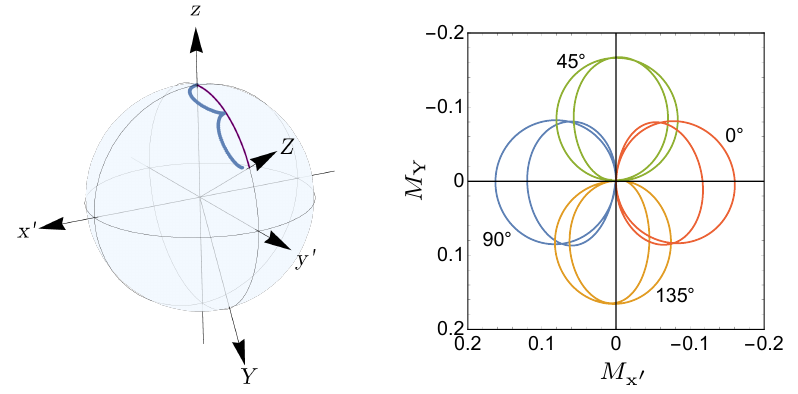} 
\caption{
Left: The rotating ($\mathrm{x', y',  z}$) and doubly-rotating ($\mathrm{x', Y, Z}$) frames with
simulations starting from unit equilibrium magnetization for a CP field (thin purple line) and linear field (thick blue line) with $B_1 =  B_\omega / \nu_\pi$ (here $\nu_\pi =3$), 
 $\delta B=0$,
$\phir=0^\circ$, and $\phis=\phid=90^\circ$.  The trajectories are shown for  a single rf period.  
The terminus of the trajectory for the linear field has a noticeable non-zero $M_{x'}$ component.
Right: 
Components of the transverse magnetization in the doubly-rotating frame shown for simulations with 
$\delta B_{\rm p} (\phis)$ of Eqs.~8 and~9 for  a variety of start phases ($\phis=\phid$, indicated on the graph) for a single period of a linear rf field. All other parameters are the same as above.
Under these conditions,   $M_{x'} \approx M_Y \approx 0$ every half-period from the start of the pulse. Deviations from perfect nullification (not visible at this scale for this example) can be eliminated by a small change in $B_1$ and a higher order correction in $\delta B_{\rm p}$, as discussed in Sec.~3.2. 
}
\label{DR_frame}
\end{figure}

We now transform Eq.~\ref{B1_rot_appendix} from the rotating frame defined by axes ($\bm{\hat{x}'},  \bm{\hat{y}'},  \bm{\hat{z}}$)   to a doubly-rotating frame defined by axes ($\bm{\hat{x}'},  \bm{\hat{Y}},  \bm{\hat{Z}}$) 
as shown in Fig.~\ref{DR_frame},  where $ \bm{\hat{Y}}=\bm{\hat{y}'} $ and
$ \bm{\hat{Z}}=\bm{\hat{z}} $ until $t=t_d$. 
The result is
\begin{eqnarray}
\bm{\mathcal{B}}_1^{++}(t)
\!\!\! 	& = & \!\!\!  
\tfrac{1}{2}B_1  \left[ (1+\cos 2\omega t)\ \bm{\hat{x}'}  \right. \\ 
\!\!\! 	& + & \!\!\!  \left.    \sin2\omega t\ (\cos\omega_1 \tp \ \bm{\hat{Y}}+ \sin\omega_1 \tp  \  \bm{\hat{Z}} ) \right]  
\label{B1_drot_appendix} 
\end{eqnarray}
where the pertinent transformations between frames are
\begin{align}
\bm{\hat{z}} &  =   \cos\omega_{1}\tp  \  \bm{\hat{Z}}  -   \sin\omega_{1}\tp \  \bm{\hat{Y}}  \label{Eq4}\\
\bm{\hat{y}'}  &  =   \cos\omega_{1}\tp  \ \bm{\hat{Y}} +   \sin\omega_{1}\tp  \ \bm{\hat{Z}}  \, ,\label{Eq5} \, 
\end{align}
with $\omega_{1} = |\gamma| B_1/2=\omega/(2\nu_\pi)$ and $ \tp=\left(  t-\td\right)$ being the time from the start of the pulse.
The magnetization $\bm{M}$ would remain constant along $\bm{\hat{Z}}$  in this frame  for a
resonant CP field, but cyclic deviations are induced by  the CR component of the linear  rf field. 
This is shown in Fig.~\ref{DR_frame} using example simulations of the type discussed in Sec.~3.2.

 The full expressions for the Bloch equation in the doubly-rotating frame can be shown to be
\begin{align}
\dot{M}_{x'}  &  =\omega_{1} \sin2\omega t\  \left[   \sin
\omega_{1}\tp  \ M_Y - \cos
\omega_{1}\tp  \ M_Z \right] \nonumber  \\
& -\delta 
 \left[ \cos\omega_{1}\tp  \ M_Y +\sin\omega_{1}\tp  \ M_Z \right] \label{Mdotx}\\
\dot{M}_Y  &  =\omega_{1}\left[  \cos2\omega t\ M_Z%
-\sin2\omega t\ \sin\omega_{1} \tp  \ M_{x'}\right]  \nonumber  \\
&+\delta\ \cos\omega_{1}\tp  \ M_x' \label{MdotY} \\
\dot{M}_Z  &  =\omega_{1}\left[  \sin2\omega t\ \cos\omega
_{1} \tp  \ M_{x'} -\cos2\omega
t\ M_Y \right]  \nonumber  \\
& +\delta\ \sin\omega_{1} \tp
\ M_{x'} \, ,\label{Mdotz}
\end{align}
where for notational ease the field shift is written as $\delta \equiv \left|\gamma\right| \delta B_{\rm p}$ in units of angular frequency.
Consistently solving the Bloch equation 
by integration requires some care, even for the lowest
orders in $\omega_{1}/\omega$ (which scales as $1/\nu_\pi$) and $\delta/\omega_{1}$.
 Equation~\ref{Mdotz} does not contribute, for example,
since $M_Z$ deviates from 1 only to second order. 

We start by integrating an approximation of Eq.~\ref{MdotY} in which we set $M_Z=1$ 
and $M_{x'}=0.$ The corresponding first order solution
for $M_Y$ is
\begin{equation}
M_Y^{(1)}(t)=\frac{\omega_{1}}{2\omega}\left(  \sin2\omega t -\sin
2\phis \right)  \, , \label{solY}%
\end{equation}
which we note vanishes every half-period from the beginning of the
pulse at $\td$ (see Fig.~\ref{DR_frame}) and in particular at the end of
 pulses with $\nu_\pi = n/2$. Equation~\ref{solY} is then plugged into the first term
of Eq.~\ref{Mdotx} together with $M_Z=1$, but is neglected in the second term.
This yields the basis for the lowest-order solution for $M_{x'}(t)$:
\begin{align}
M_{x'}^{(1)}(t)  &  = \int\nolimits_{t_d}^{t}  \!\!\! dt' \left\{
\frac{\omega_{1}^2}{2\omega}\left[  \sin^{2}2\omega t'    - \sin2\omega t' \ \sin
2\phis \right]  \ \sin\omega_{1}\tp'   \right. \nonumber \\
&- \omega_{1} \left. \sin2\omega t' \ \cos\omega_{1}\tp'    - \delta  \sin\omega_{1} \tp' \right\}  \, ,
\label{solx}
\end{align}
where $ \tp' =\left(t'-\td\right)$.

We now consider making a $\pi$-pulse in $\nu_\pi$ periods of the rf field by setting the upper limit in 
Eq.~\ref{solx} 
to $t = \td +  \nu_\pi T$ and performing the integration. At the end of the pulse, $\bm{\hat{Y}}=-\bm{\hat{y'}}$ and $\bm{\hat{Z}}=-\bm{\hat{z}}$ as expected from Eqs.~\ref{Eq4} and~\ref{Eq5} and Fig.~\ref{DR_frame}.

For the special case of $\nu_\pi = n/2$, i.e., a pulse duration equal to an integer multiple of the rf half-period, one finds to lowest orders in $\omega_{1}/\omega$ and $\delta/\omega_{1}$  that
\begin{equation}
M_{x'}^{(1)} \approx \frac{\omega_{1}}{4\omega} \left( 1 - 2\cos 2\phis - \frac{4 \omega \delta}{\omega_1^2}  \right)
\left(  1-\cos\pi n\omega_{1}%
/\omega \right) \, , \label{solX}
\end{equation}
As a result, $M_{x'} \approx 0$ at the end of the $\pi$-pulse if one chooses
\begin{equation}
\delta 
=  \left( 1-2\cos2\phis \right) \times \omega \left(  \frac{\omega_{1}}{2\omega}\right)^{2} 
=  \beta(\phis) \, \left( \frac{\omega}{16\nu_\pi^2} \right) \, ,
\label{BS_derived}
\end{equation}
as per Eqs.~8 and~9.  
Another important feature of Eq.~\ref{solX} 
is that one can factor out the term $(1-\cos\pi n \omega_{1}/\omega)$, which was purposely left unevaluated here to highlight this fact.
This in turn implies that with the correct field shift, $M_{x'} \approx M_{Y} \approx 0$ every half-period from the start of the pulse and $M$ is indeed on the expected CP trajectory at these points as discussed in Sec.~3.
This also implies that at the end of a $\pi$-pulse with $\nu_\pi = n/2$, a terminus at the south pole is achieved to lowest significant order with $B_1$ being the nominal RWA rf amplitude, as assumed throughout this analysis. 

The above procedure can be repeated for a general $\pi$-pulse characterized by any value of $\nu_\pi$, and the corresponding value of $\beta$ can be extracted.  Letting $\nu_\pi = n/2 + \delta \nu_\pi $, where $\delta \nu_\pi$ is the difference in pulse duration from the case 
$\nu_\pi = n/2$ and ranges from $-1/4$ to $1/4$ as shown in Fig.~9,  one finds
\begin{eqnarray}
\beta(\phis,\phie) & = &1 -\cos2\phis -\cos2\phie \\ 
& = &1 -\cos2\phis -\cos(2\phis +4\pi\delta \nu_\pi)
\label{eq:globalbeta}
\end{eqnarray}
where $\phie = \phis + 2 \pi \, (n/2 +\delta \nu_\pi)$ is the phase at the end of the pulse. 
 A summary of some specific results related to Figs.~8 and~9 are
\begin{equation}
\beta(\phis)=
\begin{cases}
   1-2\cos2\phis, & \text{for } \nu_\pi= \tfrac{n}{2} \\
   1-\cos2\phis  \pm \sin2\phis , & \text{for } \nu_\pi= \tfrac{n}{2} \pm \tfrac{1}{8} \\
    1, & \text{for } \nu_\pi= \tfrac{n}{2} \pm \tfrac{1}{4} 
  \end{cases}
\label{BS_GTderived}
\end{equation}

The condition of Eq.~\ref{eq:globalbeta} only  ensures that $M_{x'} \approx 0$ at the end of the $\pi$-pulse.  From Eq.~\ref{solY} the general value of $M_Y$ at this point is
\begin{equation}
M_Y \approx \frac{1}{4\nu_\pi}\left(  \sin(2\phis + 4\pi \, \delta \nu_\pi)  - \sin 2\phis \right)  \, .
\label{MYafterPi}
\end{equation}
When $ \delta \nu_\pi \neq 0$, $M_Y$ will be zero for  specific values of $\phis$ only.  In general, non-zero values of $M_Y$ arising from Eq.~\ref{MYafterPi} are an indication that $B_1$ needs to modified from the value $B_\omega / \nu_\pi$ to achieve an accurate $\pi$-pulse to lowest order. 

Alternative analytic calculations of trajectories on the Bloch sphere have been made
based on F. Ansbacher's semiclassical work on approximate equations of motion for
atomic spins in a magnetic field with static and oscillatory components~\cite{Ansbacher1973}.
The results obtained for rectangular pulses with this approach, using the coordinate transformation explicitly
derived (to the lowest perturbative order in $B_{1}/B$) for a nearly resonant transverse linear field with constant
amplitude $B_{1}$,  are in good agreement with those presented here and in the main text.

\renewcommand\thefigure{B.\arabic{figure}}    
\renewcommand{\theequation}{B.\arabic{equation}}
\setcounter{figure}{0}    
\setcounter{equation}{0}

\section{Hyperbolic fitting function for near-\texorpdfstring{\bm{$\pi$}}{Pi} pulses}
\label{sec:hyperbolas}

Here we consider circular trajectories on the Bloch sphere in the RWA for rectangular rf pulses of fixed duration. We define $f_1^{(0)}=\left(\left|\gamma\right|/2\pi \right) \times B_\omega T/(2t_\pi )$ to be the nutation frequency associated with the amplitude $B_\omega / \nu_\pi $ of a linear rf field pulse of duration $t_\pi = \nu_\pi  T$ (Eq.~7) needed to achieve an exact $\pi$-pulse at resonance, i.e. for $\Delta f=0$ as per the notation of Sec.~\ref{sec:DataAquisitionSM}. For an rf amplitude 
$B_1 = (1+b_1) \times (B_\omega / \nu_\pi) $ applied out of resonance, 
the rf-driven nutation occurs around an effective field tilted by a polar angle $\Psi$ from 
the positive $z$-axis:
\begin{equation}
\Psi=\arctan \left[\frac{1+b_1}{\left(\Delta f/f_1^{(0)}\right)}\right] \, ,
\label{Rabi1}
\end{equation}
where $|b_1| \ll 1$ for near-$\pi$ pulses.
The rotation angle $\psi_\mathrm{e}$=2$\pi f_\mathrm{e}\tau$ around the tilted nutation axis results from the
effective nutation frequency $f_\mathrm{e}$ and the pulse duration $t_\pi$: 
\begin{align}
f_{e}^2 & = \left(f_1^{(0)}\right)^2\:\left[\left(1+b_1 \right)^{2}+\left(\Delta f/f_1^{(0)} \right)^{2}\right]\\
\psi_\mathrm{e} & = \pi f_e / f_1^{(0)}.
\label{Rabi2}
\end{align}

At the end of the pulse, the $z$-component of the unit vector $\bm M$ initially along 
 the $+z$-axis is
\begin{equation}
M_{z}  =\cos^{2}\Psi+\sin^{2}\Psi\cos\psi_\mathrm{e}
\label{Rabi3}
\end{equation}
and the flip angle $\theta$ is
\begin{equation}
\theta=\arccos(M_{z}).
\label{Rabi4}
\end{equation}

For small deviations from an exact $\pi$-pulse, Eq~\ref{Rabi4} can be expanded in powers of the small control parameters. To the lowest significant orders in $b_1$ and $\Delta f/f_1^{(0)},$ the equation of an hyperbola is derived:
\begin{equation}
\left(\pi - \theta\right)^{2}=4 \left(\Delta f/f_1^{(0)} \right)^{2}+b_1^{2} \pi^{2}+...
\label{Rabi5}
\end{equation}

For rectangular pulses lasting a half-integer number of periods ($2\nu _\pi$ being an integer number), theory and numerical simulations show that the tipped unit magnetization lies on the RWA trajectory 
obtained for a modified value of  $\Delta f$, as demonstrated in Fig.~4. 
Equation \ref{Rabi5} therefore justifies the use of 
the following 
hyperbolic fitting function for sets of experimental data such as displayed in Figs.~16b and 16e where $\pi - \theta $ is plotted versus the rf amplitude parameter $V_\mathrm{m}$ at fixed detuning: 
\begin{equation}
y\left(V_\mathrm{m}\right)
= \left[ \left(2 \frac {\Delta f_\mathrm{fit}} {f /2\nu _\pi } \right)^{2}+\left( s_\mathrm{fit} \pi \frac {V_\mathrm{m}-V_\mathrm{fit}} {V_\mathrm{fit}}\right)^{2} \right]^{1/2}.
\label{fitfunc}
\end{equation}
The nominal nutation frequency $f_1^{(0)}$ was replaced here with  its equivalent value 
$f /2\nu _\pi $, where $f =\omega/2\pi$ is the rf frequency.  The fitting function involves three fit parameters: $\Delta f_\mathrm{fit}$ is the effective frequency shift which would yield the same data in the RWA, $s_\mathrm{fit}$ is a slope scaling parameter expected to be 1, and $V_\mathrm{fit}$ is the amplitude parameter for which $\pi - \theta $ is minimal.

\renewcommand\thefigure{C.\arabic{figure}}    
\renewcommand{\theequation}{C.\arabic{equation}}
\renewcommand{\thetable}{C.\arabic{table}}
\setcounter{figure}{0}    
\setcounter{equation}{0}
\setcounter{table}{0}
\section{RF coils}
\label{sec:RFcoil}

\begin{figure}[htb]
\centering
\includegraphics[trim=0 0 0 0, clip=true, width=0.8\columnwidth]{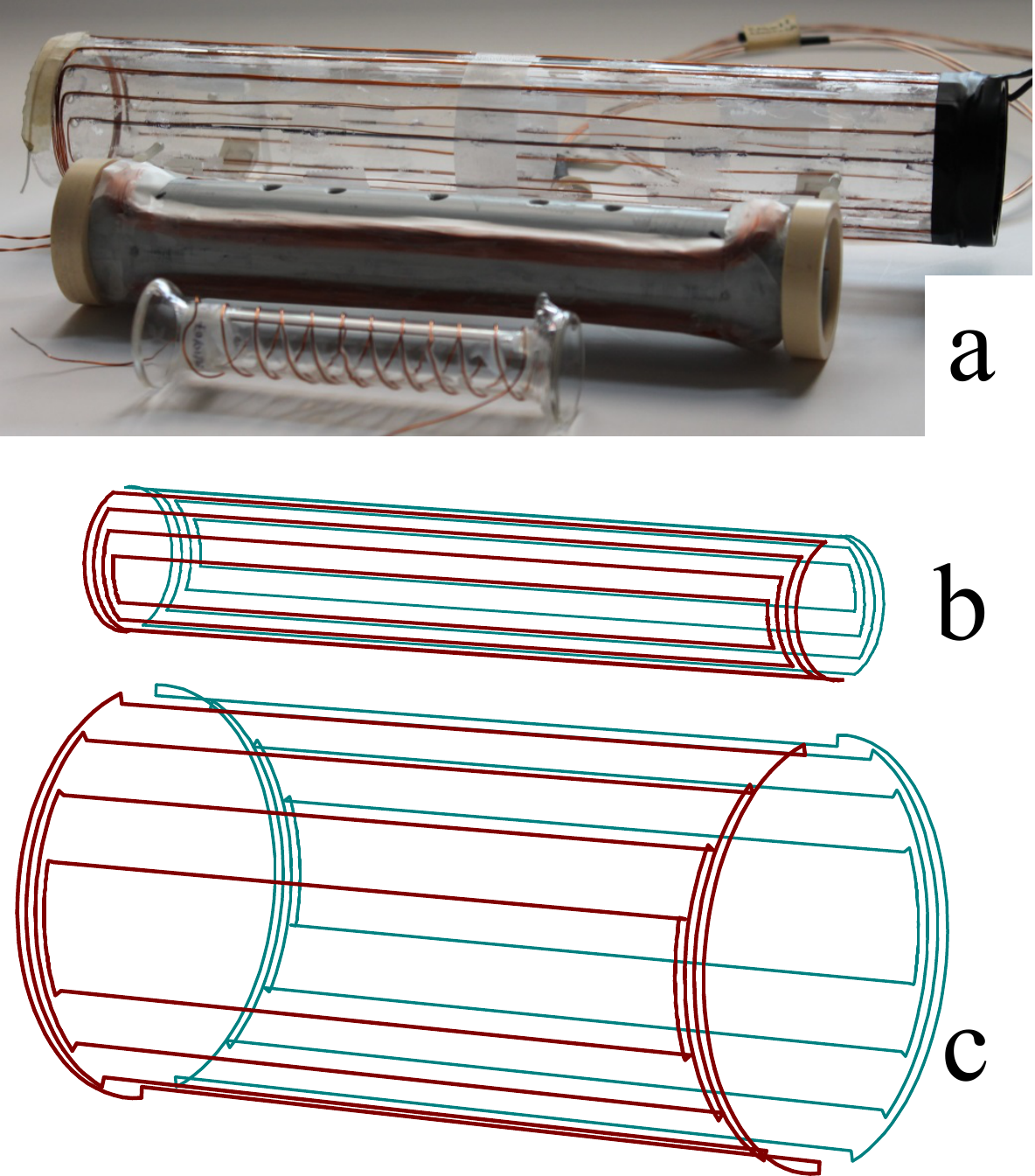}
\caption{
Photograph and design of the rf coils. a: The Sm$_{\mathrm{lowZ}}$ and PU coils  and $^3$He cell are displayed side-by-side for clarity (they are normally all co-axial with the main static field). The electrodes wound on the cell can be seen. b and c: Wire-sketches of the four saddle-coil elements of the small (b) and large (c) transmit coils. 
}
\label{fig-exp2}
\end{figure}

The four different rf coils  employed in this work (Sm, Sm$_{\mathrm{lowZ}}$, Lg, and PU) are all cylindrical in design, utilizing appropriately distributed saddle-coil elements (e.g., Fig. \ref{fig-exp2})
in order to approximate the ideal sine-phi surface current distribution (e.g., Appendix~D of Ref.~\cite{Bidinosti2005}) and provide a uniform, transverse, internal field.  The design and electromagnetic parameters of all coils are summarized in Table~\ref{tab:Coils}.  
The photograph in Fig. \ref{fig-exp2} shows the Sm$_{\mathrm{lowZ}}$ and PU coils as well as the sealed $^3$He cell. 
The 12-cm long cell is a Pyrex tube (15~mm i.d.) closed at each end with a 30-mm diameter flat Pyrex window, resulting in a \textit{bone-shaped} design suitable for high-pressure MEOP~\cite{Gentile}. 

The Sm, Sm$_{\mathrm{lowZ}},$ and Lg coils all comprise four saddle-coil elements of average length 28~cm, whose number of turns and current arcs  were determined by numerical optimization of field homogeneity.  The Sm$_{\mathrm{lf}}$ and Sm$_{\mathrm{hf}}$ coils are identical in design (Fig.~\ref{fig-exp2}b), with the latter having one third the number of turns on each element to reduce its inductance for  operation at higher frequency ($f_\mathrm{high}=83.682$~kHz versus  $f_\mathrm{low}=25.7$~kHz).
They were wound using 0.4-mm-diameter enamelled copper wire on the outer surfaces of 3-mm-thick, 5-cm-o.d., 29-cm-long PMMA tubes.
The Lg coil was wound on a 3-mm-thick, 15-cm-o.d., 29-cm-long PMMA tube using the same wire. The axial current rungs lie on the inner surface of the tube and the end segments on the outer surface (Fig.~\ref{fig-exp2}c).  
Whenever used at $f_{\mathrm{high}}$, Lg was tuned with a series 9.5-nF capacitor, giving a quality factor of about 15 and reducing its impedance from 174~$\Omega$ to 12~$\Omega$, which made it better matched to the low-impedance RF amplifier. 

\begin{table}[H]
\centering%
\begin{tabular}[c]{l | llll}
& Sm & Sm$_{\mathrm{lowZ}}$ & Lg & PU\\\hline
$n_1$ (turns) & 3 & 1 & 2 & 100\\
$n_2$ & 6 & 2 & 4\\
$n_3$ & 9 & 3& 6\\
$n_4$ & 9  & 3 & 6\\
$\varphi_1$ (degrees) & 19.6 & 19.6 & 19.1 & 60\\
$\varphi_2$ & 39.2 & 39.2 & 39.2\\
$\varphi_3$ & 59.7 & 59.7 & 59.8\\
$\varphi_4$ & 81.5 & 81.5 & 82.0\\
$B/I$ (mT/A) & 0.671 & 0.224 & 0.170 & 3.72\\
$L$ (mH) & 0.49 & 0.060 & 0.33 & 5.1\\
$2\pi Lf_{\mathrm{low}}$ ($\Omega $) & 79 & 9.7 & 53 & 824\\
$2\pi Lf_{\mathrm{high}}$ ($\Omega $) & 258 & 31.5 & 174 & 2700\\
$R_{\mathrm{dc}}$ ($\Omega $) & 5.5 & 2.2 & 5.0 & 12.6\\
\hline
$\tau_\mathrm{H}$ ($\mu$s) & 51 & 11 & 36 & \\
$\tau_\mathrm{T}$ ($\mu$s) &  8.7 & 1.1  & 5.9 & \\
\end{tabular}

\caption{
Table of rf coil parameters: $n_j$ and $\varphi_j$ are the number of turns and current arc half-angles~\cite{Bidinosti2005} for each half of the $j$-th saddle-coil element; $B/I$ is the field-per-unit-current or efficiency at the coil iso-centre; $L$ is the inductance; and $R_{\mathrm{dc}}$ is the dc resistance.  
Bottom: Time constants of the untuned coil circuits. For the home amplifier, $\tau_\mathrm{H}$ was determined from recorded transients.
For the Tomco amplifier,  $\tau_\mathrm{T}$ was computed as $L$ divided by the total circuit resistance. 
} 

\label{tab:Coils}
\end{table}

The PU coil is a single 18-cm-long saddle coil wound on a 1.75-mm-thick, 3.4-cm-o.d.\ PVC tube. For all experiments it was arranged with its rf-field direction perpendicular to that of the accompanying transmit coil to minimize cross-talk and transient saturation of the detection channel following rf  pulses. PU was tuned with a parallel capacitor. The open-loop quality factor of the tank circuit $Q_{\mathrm{ol}}$=25 at $f_{\mathrm{low}}$ (50 at $f_{\mathrm{high}}$) was reduced to $Q_{\mathrm{FB}}$=3 at $f_{\mathrm{low}}$ (12 at $f_{\mathrm{high}}$) using a suitable amount of negative feedback of the amplified detected signal to the tank circuit  \cite{Baudin11,Kuzmin2020}.  This was done in order to provide a flat, broad-band response, without incurring any SNR penalty, for experiments where frequency was varied.
A pair of  crossed diodes (1N5849) was used to clip cross-talk signals and a reed relay with two normally closed contacts (DIP05-2A72-21L, MEDER Electronic) was used in some experiments to open connections of the coil to the tuning capacitor and detection circuit during pulses, thereby preventing any rf field perturbation by induced currents flowing in PU. Noise from the rf amplifier was blocked between pulses using crossed diodes (1N4004) or a mechanical relay (IM23TS, TE Connectivity) in series with the coil.

\balance
\putbib
\end{bibunit}

\end{document}